\documentclass[aps,prl,amsmath,amssymb,superscriptaddress,reprint]{revtex4-2}

\usepackage{graphicx}
\usepackage{dcolumn}
\usepackage{bm}
\usepackage{multirow}
\usepackage[utf8]{inputenc}
\usepackage[T1]{fontenc}
\usepackage{mathptmx}
\usepackage{etoolbox}
\usepackage[version=4]{mhchem}
\usepackage{cleveref}
\usepackage{float}
\usepackage{url}
\usepackage{xcolor}

\usepackage{amsmath}
\usepackage[version=4]{mhchem}

\newcommand{\applabel}[2]{\expandafter\def\csname @currentlabel\endcsname{#1}\label{#2}}

\begin{document}

\title{Adaptive Slater Koster Parameters:\\ Crossing Oxidation States with Density Functional Tight Binding}

\author{Yihua Song}
    \affiliation{Fritz Haber Institute of the Max Planck Society, Faradayweg 4-6, 14195 Berlin, Germany}

\author{Artem Samtsevych}
    \affiliation{Fritz Haber Institute of the Max Planck Society, Faradayweg 4-6, 14195 Berlin, Germany}

\author{Anton Beiersdorfer}
    \affiliation{Fritz Haber Institute of the Max Planck Society, Faradayweg 4-6, 14195 Berlin, Germany}

\author{Tobias Melson}
    \affiliation{Max Planck Computing and Data Facility, Gießenbachstraße 2, 85748 Garching b. München, Germany}

\author{Christoph Scheurer}
    \affiliation{Fritz Haber Institute of the Max Planck Society, Faradayweg 4-6, 14195 Berlin, Germany}
    
\author{Karsten Reuter}
    \affiliation{Fritz Haber Institute of the Max Planck Society, Faradayweg 4-6, 14195 Berlin, Germany}

\author{Chiara Panosetti}
    \email{panosetti@fhi.mpg.de}
    \affiliation{Fritz Haber Institute of the Max Planck Society, Faradayweg 4-6, 14195 Berlin, Germany}

\date{\today}

\begin{abstract}

We propose to adapt the confined pseudo-atomic orbitals underpinning the Slater-Koster (SK) tables in density functional tight binding (DFTB) to local atomic environments. Assigning optimal SK parameters to metal atoms in different oxidation states improves the electronic structure of a partially oxidized Ni surface and, with adapted repulsion, Li insertion energetics in graphite. Their smooth, interpretable response to chemical and spatial environment, independently observed for Ni and Co oxides, enables a simple, physically informed machine-learning scheme, here applied to Ni–O binaries.
\end{abstract}

\maketitle

As a stationary point approximation to density-functional theory (DFT), Density Functional Tight Binding (DFTB)~\citep{Elstner1998} builds on ideas originating from the work of Slater and Koster~\citep{SlaterKoster} and the Foulkes-Haydock interpretation of tight binding~\citep{PhysRevB.39.12520}, developed by Seifert, Porezag and Elstner~\citep{Seifert1996,Porezag1994,Elstner1998} into the simple total energy formula: 
\begin{equation} \label{eq:total_e}
E_{\mathrm{tot}} \;=\; E_{\mathrm{el}} + E_{\mathrm{rep}} \quad .
\end{equation}
Here, $E_{\mathrm{el}}$ is a parameterized electronic energy and the so-called repulsive energy $E_{\mathrm{rep}}$ represents a correction including the short-range nuclear repulsion as well as all the contributions not captured by the approximated electronic energy. In successively more sophisticated formulations, $E_{\mathrm{el}}$ takes the form of increasing orders of expansion around a reference non-interacting density. In the zeroth-order DFTB0 (often also called DFTB1)~\citep{Seifert1996,Porezag1994}, $E_{\mathrm{el}}$ is the so-called band-structure energy $E_{\mathrm{BS}}$, where the non-self-consistent interaction is pre-calculated in a two-center approximation by solving a set of generalized eigenvalue equations using a minimal atomic-centered valence basis set under a parametric confinement, and tabulated in so called Slater-Koster (SK) integrals. This allows to access the electronic structure without painfully solving expensive self-consistent field equations at runtime. In the second- and third-order formulations (DFTB2/3~\cite{Elstner1998, DFTB3}), $E_{\mathrm{BS}}$ is augmented with a charge-dependent Coulomb term $E_{\mathrm{coul}}$. These formulations, also referred to as self-consistent-charge DFTB (SCC-DFTB), are nowadays the most widely used, with freely available simulation packages such as {\tt dftb+}~\citep{Hourahine2020}. For a detailed derivation, we defer to existing literature~\citep{Seifert1996,Porezag1994,Elstner1998, DFTB3}. 
With decades of development, DFTB has emerged as a leading semi-empirical method in large-scale theoretical research as it offers an excellent compromise between cost-effectiveness and accessible electronic structure accuracy~\cite{dral2019semiempirical,dral2024modern}, enabling modelling of systems with thousands of atoms while preserving the description of essential quantum interactions, such as electron correlations and charge fluctuations~\cite{Elstner1998}. Consequently, DFTB has demonstrated effectivity and versatility in applications ranging from molecular to materials simulations. 

However, despite higher-order expansions, intrinsic limitations persist due to approximations and parameterization challenges~\cite{Hourahine2020}. For instance, previous studies highlighted the inadequacy of a single repulsive fit in both organic~\cite{Gaus2013} and solid state~\cite{AmmothumKandy2021} systems. While extended tuning~\cite{Hellstrom2013,Fihey2015} and machine learning (ML) or many-body approaches~\cite{Panosetti2020,stoehr2020accurate,AmmothumKandy2021,polynominal_10.1063/5.0141616} provide partial solutions, repulsive fitting only corrects the total energy and forces, not the electronic structure. The minimal basis set and parametric nature of the precalculated interaction tables struggle in complex local environments~\cite{Chiara2023}, especially in systems with rich redox chemistry and massive charge transfer, like nickel oxides in various coordination types~\cite{Nickel-DFTB3}. 
Analogously to the repulsion, a single, per-element fit of the electronic interaction—neglecting local environment subtleties—limits DFTB’s applicability to complex materials, especially those requiring accurate description of atomic charges and electronic features near the Fermi level. 
Recent efforts have begun to address this limitation through ML. Differentiable programming frameworks have enabled the optimization of DFTB electronic parameters, including environment-dependent compression radii, with demonstrated success for molecular systems and periodic solids with point defects~\cite{Fan2022,McSloy2023,Sun2023}. Direct learning of tight-binding Hamiltonians~\cite{Hu2023,Gu2024} and many-body corrections~\cite{Burrill2025} have also been explored. 
Building on this progress, we go beyond black-box learning environment-dependent parameters to establish the physical basis for confinement adaptation when distinct oxidation environments of the same element coexist.

In this Letter, we show that in such systems, optimal confinement parameters respond systematically, decomposably and interpretably to both spatial and chemical factors, establishing the physical foundation for a site-resolved adaptive DFTB scheme. Starting with a simple rationalization in terms of discrete atomic types
and using a mixed-oxidation prototype, we demonstrate that locally appropriate confinements recover the reference ordering of site-resolved Ni 3d populations and improve the electronic structure near the Fermi level significantly.
Scanning tunneling microscopy (STM) simulations of a stepped, partially oxidized Ni surface and exemplary reaction paths for lithium insertion into graphite illustrate its use for both electronic structure and energetics. Crucially, the observed smoothness of SK parameters across chemical environments not only ensures physical interpretability but allows us to introduce a simple ML scheme for continuous adaptation beyond the discrete picture, offering a data-efficient route to improved electronic structure in large-scale atomistic simulations.

To understand why the DFTB parameters intrinsically need local environment adaptation, we shall first briefly discuss some elements of DFTB parameterization. The two interaction terms in~\cref{eq:total_e} are routinely parameterized separately. The repulsion is parameterized by minimizing the deviation in force residues between DFT and repulsion-less DFTB. For long considered challenging, repulsive fitting is now largely facilitated by ML~\cite{Panosetti2020,stoehr2020accurate,polynominal_10.1063/5.0141616}. Hence, we turn our attention to the electronic parameterization, which is often oversimplified. The generalized Kohn-Sham equation \citep{PhysRev.140.A1133} in a tight-binding ansatz can be written as follows:
\begin{flalign}
\label{generalized eigen}
&\sum_\nu^M c_{\nu i} (H_{\mu \nu} - \epsilon_i S_{\mu \nu}) = 0 , \quad \forall \mu,i ; \\ 
\begin{split}\label{SK integral}
H_{\mu \nu} &= \langle \varphi_\mu \mid\hat{H}_0\mid \varphi_\nu \rangle +\frac{1}{2} S_{\mu \nu}\sum^N_{\xi} (\gamma_{\alpha\xi}+\gamma_{\beta\xi})\Delta q_\xi   \\
&= H^0_{\mu \nu} +H^1_{\mu \nu}, S_{\mu \nu} \;=\; \langle\varphi_\mu\mid\varphi_\nu\rangle , \quad  \forall \mu\in\alpha, \nu\in\beta
\end{split}
\end{flalign}
where $c_{\nu i}$ are the expansion coefficients of state $i$ with eigenvalue $\epsilon_i$ in a basis of $M$ non-orthogonal pseudo-atomic orbitals $\{\varphi_\mu\}$, with $\mu$ and $\nu$ centered on atoms $\alpha$ and $\beta$, and the sum in~\cref{SK integral} runs over all $N$ atoms. The two-center matrix elements $H^0_{\mu \nu}$ of the reference Hamiltonian $\hat{H}_0$ (at the superposition of neutral atomic densities) and the overlap $S_{\mu \nu}$ are pre-calculated following the Slater-Koster method~\citep{SlaterKoster} and stored in integral tables.
$H^0_{\mu \nu}$ is the core of the DFTB0 method, while the second-order correction $H^1_{\mu \nu}$ (core of DFTB2) represents the Coulomb interaction $E_{\mathrm{coul}}$~\cite{Elstner1998}, with $\gamma$ a screened effective Coulomb interaction that depends on the Hubbard parameters $U$, and $\Delta q_\xi$ the Mulliken charge fluctuation of atom \(\xi\) relative to its neutral reference.
The pseudo-atomic orbitals are obtained from DFT calculations of spherically symmetric, spin-unpolarized neutral atoms by solving:
\begin{equation}\label{eq:pseudo_atomic_orbitals}
\big( -\frac{1}{2} \nabla^2 + V_{\textrm{KS}}^{\textrm{eff}} \left[n_{\mathrm{atomic}}\right] + V_{\mathrm{conf}} \big) \varphi_{\nu}  = \epsilon^{\mathrm{atom}}_\nu \varphi_{\nu},
\end{equation}
where $\epsilon^{\mathrm{atom}}_\nu$ is the atomic orbital energy, $V_{\textrm{KS}}^{\textrm{eff}}[n_{\mathrm{atomic}}]$ the effective Kohn-Sham potential of the atomic density $n_{\mathrm{atomic}}$, and $V_{\mathrm{conf}}$ a confinement potential applied to mimic the compression of wave functions upon bonding. This confinement (along with, optionally, an additional confinement for $n_{\mathrm{atomic}}$) is tuned until selected electronic properties are reproduced, usually band structures of simple solids. In standard parameterization approaches, for a given set of atomic species, the confinement potential is optimized once (single-fit) to yield, at best, a compromise across the target chemical compositions. 
However, one could argue that the confinement potential acts as a static correction to the effective potential $V_{\textrm{KS}}^{\textrm{eff}}$ in~\cref{eq:pseudo_atomic_orbitals}, to closely mimic the localization behavior of atomic orbitals in the realistic environment. Accordingly, we propose that each $\varphi_{\nu}$ should be associated with an environment-dependent confinement potential that encodes tight-binding perturbations from the surrounding environment into the localized basis. In the End Matter (Appendix~\ref{app:dftb3}), we justify this choice over modifications to $\gamma$ (either via DFTB3 or a hypothetical adaptive variation of $U$). 

\begin{figure} \centering
    \includegraphics[width=0.9\linewidth]{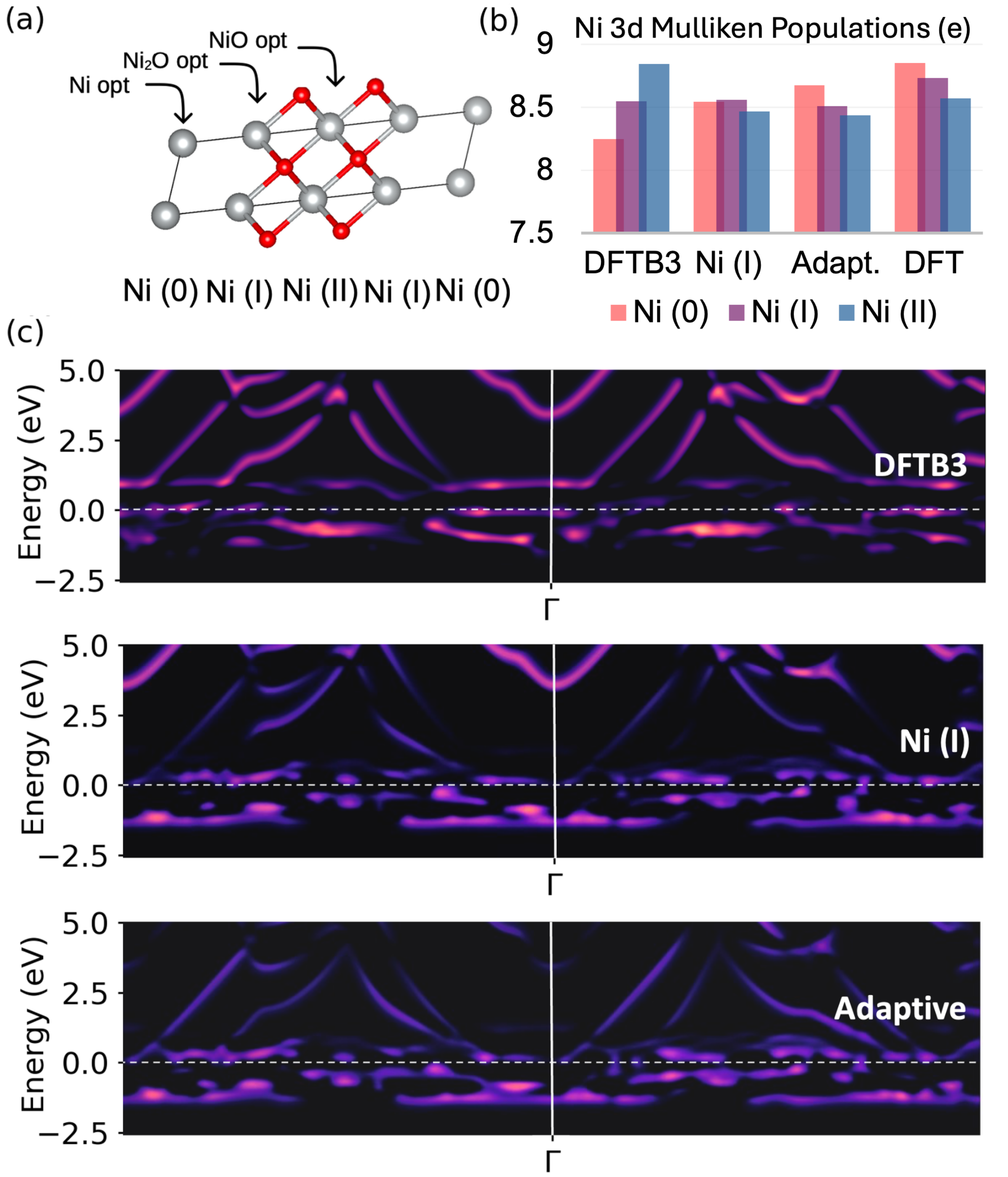}
  \caption{(a) Structure of the conceptual \ce{Ni4O2} system containing all three Ni oxidation states. (b) Mulliken population analysis for Ni $3d$ in this system. (c) Gaussian difference maps. Compared in (b) and (c) are calculations done with published DFTB3 parameters \cite{Nickel-DFTB3}, with optimal Ni(I) parameters applied globally to all Ni atoms, and with the adaptive approach using optimized parameters for each Ni oxidation state. In the difference maps, less residual coloring represents improved agreement with the DFT reference band structure (see text). The explicit band structures are reported in the SM (Figs.~\ref{SM-suppfig:DFTB3-Ni4O2} and \ref{SM-suppfig:closer-look}).
  }
    \label{fig:concept}
\end{figure}

To practically illustrate the limitation of a single-fit confinement in systems with coexisting different local environments, we choose nickel oxides with general formula \ce{NiO_x} and investigate the variation of optimal parameters across oxidation states. Aside from the technological relevance of nickel oxide for catalysis, magnetism and polaron research~\cite{charge-transfer-error,Gaus2014,NiO.TADIC20151061}, this choice is motivated by the fact that the conservation of the bulk cell shape from \ce{Ni} to \ce{NiO} isolates the oxidation effect from geometric changes. Let us first focus on bulk Ni, NiO, and an intermediate \ce{Ni2O} structure, modeled by removing one oxygen atom from the $(2\times1\times1)$ supercell of \ce{NiO}, as shown in Fig.~\ref{SM-suppfig:SI-params of NiOx}. 
Nickel atoms in these structures have formal oxidation states Ni(0), Ni(I) and Ni(II), respectively. Specific parameterizations for each are obtained using our newly developed optimization code \texttt{DSKO}~\cite{dsko}, with non-spin-polarized \ce{NiO_x} DFT-PBE~\cite{PBE.PhysRevLett.77.3865} band structures as references. This choice ensures comparability with published parameters without losing generality, as transferability to the spin-polarized case has been verified (SM, Sec.~\ref{SM-supp:spin}). 
Employed is the most used confinement potential, the power law model, defined as $V_{\textrm{conf}}(r)=({\frac{r}{r_0}})^\sigma$. Typically, the dominant parameter is the onset radius \(r_0\), whereas \(\sigma\) merely modulates the decay of the confined orbitals. As expected, the resulting optimal parameters are sensitive to local environments, with the dominant $r_{0, {\rm Ni}_{3d}}$ parameter increasing from 3.49 (Ni) to 3.87 (\ce{Ni2O}) to 4.01 (NiO) Bohr. 

It is then natural to hypothesize that, in principle, assigning optimal parameters to each atom in a complex system should yield a finer description of the electronic structure. To demonstrate the advantages of this concept in an intuitive discrete-type picture, we design a fictitious "all-types-in-one" \ce{Ni4O2} system. This system is of formal \ce{Ni2O} stoichiometry, obtained by removing two O atoms from a \ce{NiO} $(4 \times 1 \times 1)$ supercell. It simultaneously contains all three types of Ni atoms in one structure within Ni(0)-Ni(I)-O-Ni(II)-O-Ni(I) chains as shown in Fig.~\ref{fig:concept}a. Next to the actual DFT reference, we compare to two conventional DFTB single-point calculations where all Ni atoms are assigned the same SK parameters, once published DFTB3 parameters~\cite{mio-trans3d,Nickel-DFTB3} and once our optimal DFTB2 parameters for the intermediate Ni(I) oxidation state of the formal system stoichiometry. In our adaptive approach, we instead use the respective optimal parameters for the Ni atoms in the different oxidation states. As shown in Fig.~\ref{fig:concept}b, the latter approach yields indeed more accurate Mulliken charge populations. In fact, it is the only one that reproduces the relative ordering of the DFT reference charges for the three oxidation states. Also improved is the band structure, in particular around the Fermi level. Deferring a more detailed discussion to the End Matter (Appendix~\ref{app:gauss}), we demonstrate the latter improvement in Fig.~\ref{fig:concept} with Gaussian difference maps. In this scheme, we apply a Gaussian broadening to both the DFT reference and the DFTB band structure and then subtract both from each other. A perfect band reproduction would yield a fully black difference map, with residual discrepancies readily apparent as lighter colored regions. Clearly, the adaptive calculation using specifically assigned parameters reproduces the bands better than both the conventional DFTB3 with published parameters~\cite{mio-trans3d,Nickel-DFTB3} and the DFTB2 calculation globally using the Ni(I) parameters.

\begin{figure}
    \centering
        \includegraphics[width=0.9\linewidth]{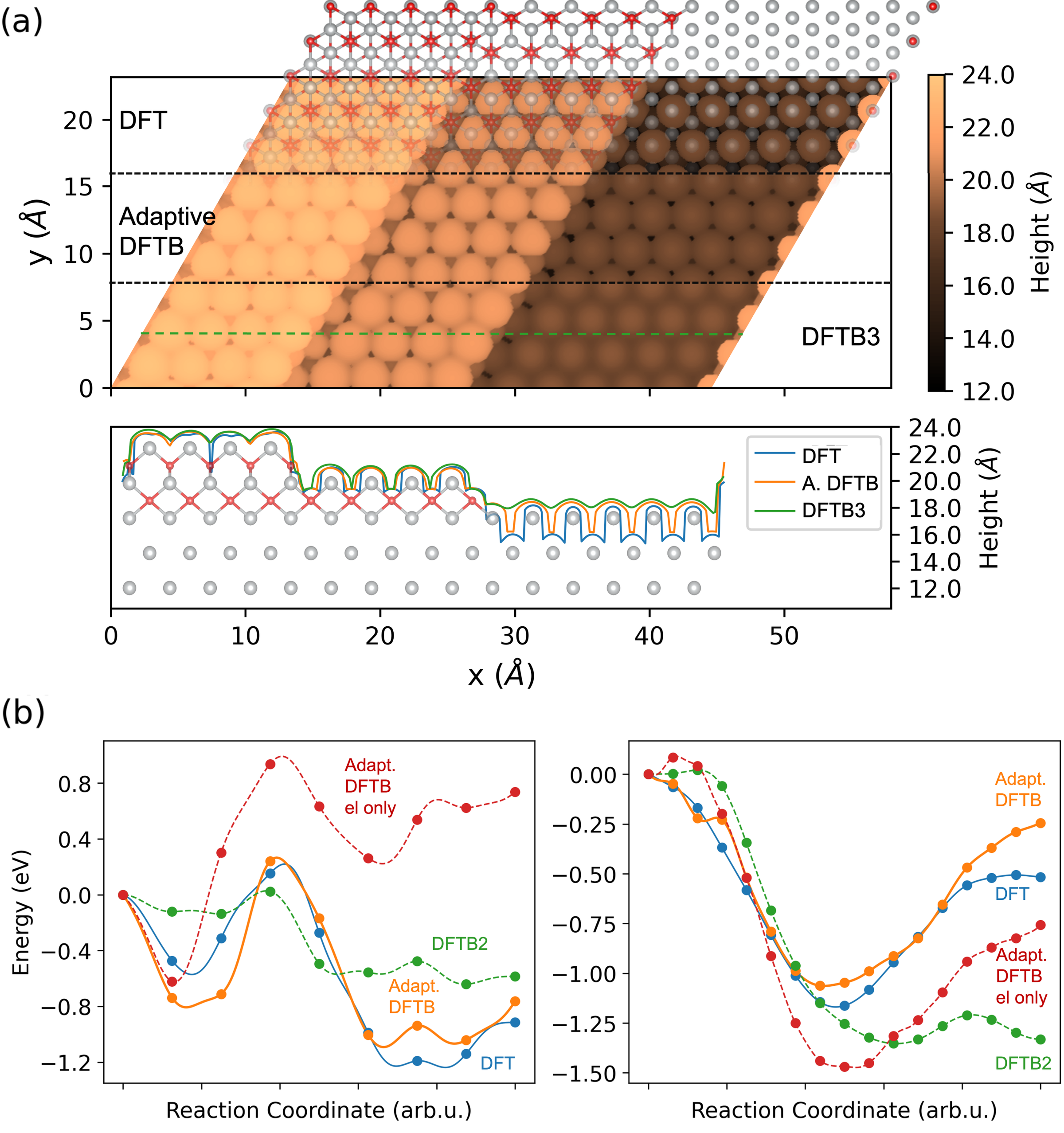}
        \caption{Comparison of adaptive DFTB calculations against the DFT reference and conventional DFTB2/3 (see text) for two application cases. (a) Tersoff-Hamann type STM simulations of a Ni(111) surface, partially covered with one and two layers of NiO surface oxide (1 V bias, 0.55 pA, corresponding to an iso-density of 4.69$\times 10^{-6} e$/{\AA}$^3$ \cite{STM_iso_density_hofer}). With both panels sharing the same $x$-axis, the upper panel shows a top view of the simulated STM image at the different levels of theory. The bottom panel shows a side view of the surface structure, overlayed with an STM line scan along the green dashed lines in (a). (b) Minimum energy paths for Li insertion into graphite through the zigzag (left) and armchair (right) terminations.}
    \label{fig:applications}
\end{figure}

To demonstrate the impact of adaptive DFTB in practical atomistic modelling applications, we present scanning tunneling microscopy (STM) simulations, which rely on a faithful electronic density around the Fermi level, and reaction paths, which require accurate total energies and forces. Tersoff-Hamann STM simulations are performed for a Ni(111) surface partially covered with one and two layers of NiO, forming a stepped surface (Fig.~\ref{fig:applications}a). While DFTB generally yields slightly more diffuse charge densities than DFT, only the adaptive approach captures the orbital compression induced by the different degrees of oxidation at each step, thus excelling in the description of localized electronic features.
For energetics, we consider Li insertion into graphite, where the inserting Li is progressively oxidized. Along nudged elastic band paths~\cite{JNSSON1998-NEB-original} obtained with published DFTB2 parameters~\cite{Panosetti2021, Annies2021}, we compare DFT and DFTB single-point energies, with discrete metallic, interfacial and intercalated Li types (SM, Sec.~\ref{SM-supp:li}). As shown in Fig.~\ref{fig:applications}b, adapting the electronic parameters alone is insufficient, and for the zigzag termination even inverts the insertion energy; only a consistently adapted repulsion yields substantially closer agreement with DFT, whereas conventional DFTB grossly misrepresents the insertion path. This discrete-type example, rather than a test of the ML scheme introduced below, thus shows that adaptive SK parameters extend to energetics, provided the repulsion is addressed accordingly.

\begin{figure}
\centering
\includegraphics[width=0.90\linewidth]{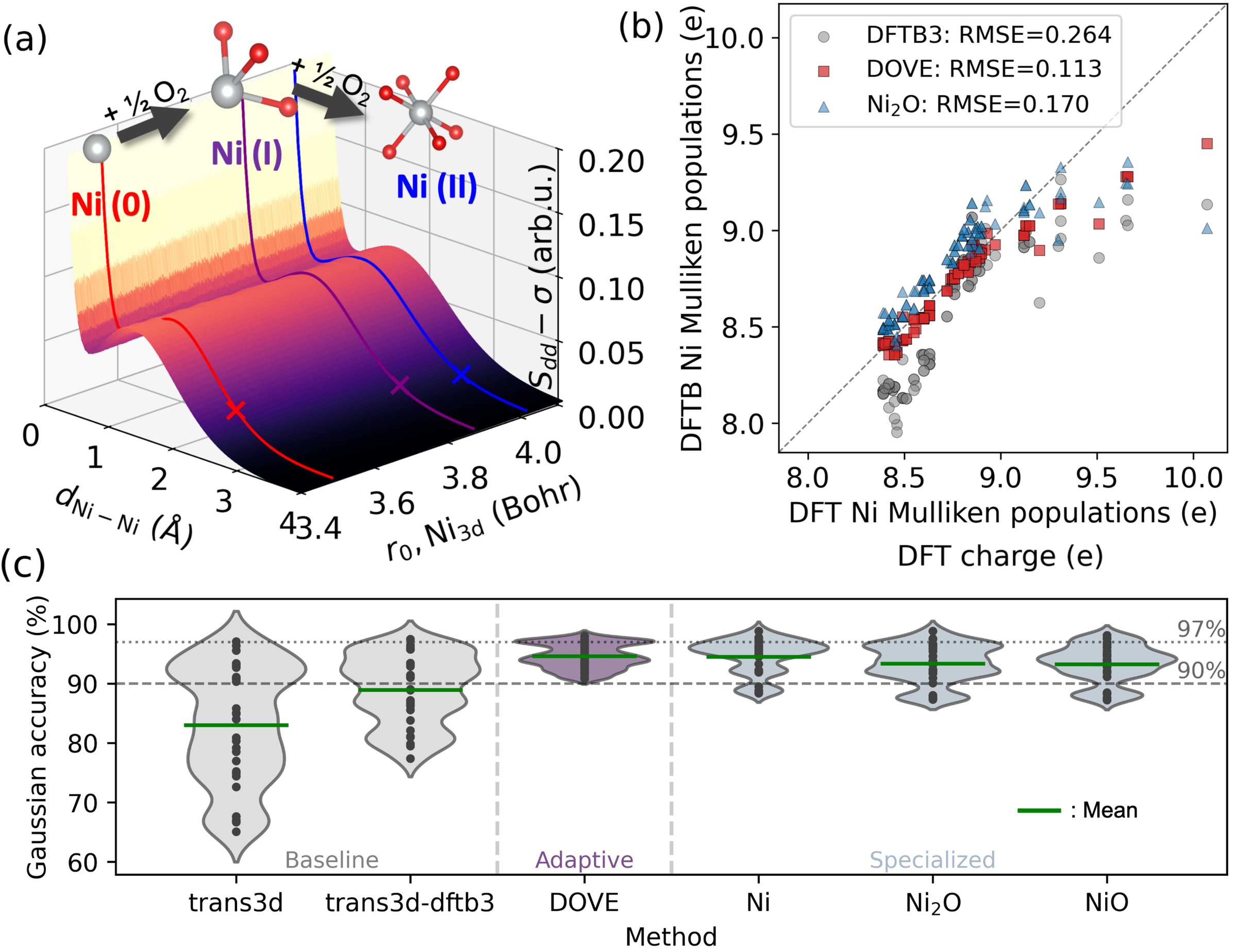}
  \caption{(a)~Plot of the exemplary $S_{dd-\sigma}$ SK integral as a function of the interatomic distance $d_{\mathrm{Ni-Ni}}$ and the onset radius $r_{0, {\rm Ni}_{3d}}$ of the confinement potential as obtained from the {\tt DSKO} optimization for \ce{Ni2O}. The variation with respect to $r_{0, {\rm Ni}_{3d}}$ is smooth, save for the noise originating from the sampling of the other, non-dominant, parameters. Optimal $r_{0, {\rm Ni}_{3d}}$ are highlighted for pure Ni (red line), \ce{Ni2O} (purple line) and NiO (blue line), and the crosses mark the specific SK parameter values taken at the $d_{\mathrm{Ni-Ni}}$ corresponding to the respective relaxed unit-cell geometry. (b)~Ni (valence) Mulliken population correlation between DFT (LOBSTER~\cite{Nelson2020}) and DFTB across the same test set in (c). (c)~Gaussian accuracy for all Ni--O binary structures from the Materials Project (see Appendix~\ref{app:gauss}).}
    \label{fig:SK_integral}
\end{figure}

While encouraging, this proof of concept is vain in practice. The discrete picture cannot be generalized simply by independently optimizing parameters for every local environment encountered in a complex simulation, as the resulting parameterization effort would quickly become impractical. The systematic variation of the dominant  $r_{0, {\rm Ni}_{3d}}$ parameter instead suggests that the environment dependence may follow an underlying physical regularity that can be rationalized and subsequently generalized.
A larger $r_0$ implies a weaker $V_\textrm{conf}$ correction to the atomic effective potential in eq. \eqref{eq:pseudo_atomic_orbitals}. This corresponds to a more diffuse orbital, whose energetic behavior is closer to a free atom Hamiltonian, and \textit{vice versa}. Here, the increasing $r_{0, {\rm Ni}_{3d}}$ suggests a more diffuse orbital in higher oxidation states, while the oxidation process, intuitively, should introduce orbital compression~\cite{Chiara2023}. This apparent contradiction is discussed in detail in the End Matter (Appendix~\ref{app:linear} and SM, Sec.~\ref{SM-supp:nio_bs}), with a systematic investigation revealing an interpretable correlation between chemical (oxidation state) and spatial (interacting distance $d$) factors. Notably, the confinement onset radius in the \ce{NiO_x} decomposes approximately as $r_0 \approx a d + b$, with $a$ capturing the spatial (distance) dependence and $b$ encoding the chemical (oxidation) shift. Crucially, this systematic and decomposable response establishes a well-defined mapping from the local chemical and spatial environment to the optimal confinement parameters. 
This regularity extends to Co--O motifs with multiple oxidation states and coordination symmetries. 
Reference-based linear regressions predict site-resolved parameters for mixed-valence spinel \ce{Co3O4} without target-specific fitting, reducing the mean absolute band-energy error by approximately 13\% relative to the better of the two uniform Co assignments (see End Matter, Appendix~\ref{app:linear}).
We thus proceed to investigate the evolution of SK integrals, the electronic coupling quantities embedded in the equations governing DFTB. Fig.~\ref{fig:SK_integral}a summarizes this evolution exemplarily for the $S_{dd-\sigma}$ SK overlap integral, which determines the Ni-Ni \textit{3d}-\textit{3d} $\sigma$ bonding interaction and is one of the quantities entering both terms of the sum in~\cref{generalized eigen} and the interacting SCC term in~\cref{SK integral}. With analog findings for other SK integrals and for Co--O (cf. End Matter), the evolution is indeed smooth. This regularity motivates a simple but consequential approach for a fully adaptive DFTB scheme beyond discrete types, where a simple linear relation no longer suffices, by physically informed ML. If the environment-to-parameter mapping is indeed well-behaved, then even a minimal regression model should be capable of predicting SK parameters for unseen configurations without material-specific re-optimization. We refer to this site-resolved prediction scheme as DOVE (DFTB Orbitals in Variable Environment). 

The critical point is not the sophistication of the regression, which is deliberately kept simple, but that the physical regularity established above suffices to support predictive parameterization across distinct stoichiometries and structures within the configurational space with only a handful of reference data. Additionally, the smoothness of the SK integrals across confinements ensures well-conditioned overlap matrices in the mixed basis, which is independently verified numerically (SM, Sec.~\ref{SM-sec:overlap_condition}).
We train Gaussian process regression models with a modified~\cite{Bartk2017, DECAF} atom-centered SOAP descriptor~\cite{SOAP_PhysRevB.87.184115} to predict the dominant onset radii $r_0$ (with $\sigma$ fixed per shell; SM, Sec.~\ref{SM-supp:gpr}), using a compact set of 60 structures containing no more than 4 atoms each, selected from a dataset generated using modular evolutionary structure search~\cite{Juan2026} to maximally capture the diversity of spatial and chemical environments in Ni--O binary bulk systems (SM, Sec.~\ref{SM-supp:training}). 
We then test the electronic structure agreement across all Ni--O binary compounds in the Materials Project~\cite{material_project_Jain2013}, spanning bulk Ni to Ni(VI), not included in the training set. As for \ce{Ni4O2}, the Ni Mulliken populations obtained with DOVE are closest to the DFT reference across the full range of oxidation states (Fig.~\ref{fig:SK_integral}b; SM, Sec.~\ref{SM-supp:mulliken}). The band structures improve likewise: DOVE yields a band-energy MAE of 0.35 eV, compared with 0.61 eV for the published DFTB3 parameters and 0.45 eV for our best single-type parameterization. The distribution of Gaussian accuracies (defined in Appendix~\ref{app:gauss}) shown in Fig.~\ref{fig:SK_integral}c demonstrates a more faithful reproduction of near-Fermi spectral features in dense band manifolds. Further validation on held-out structures from the evolutionary dataset confirms the robustness of the approach (SM, Sec.~\ref{SM-supp:heldout}).
A natural extension of the present atom-centered framework  would involve learning the SK integrals directly with two-center descriptors (e.g. Allegro~\cite{allegro-Musaelian2023}), skipping the on-the-fly reparameterization step, which, albeit accelerated (SM, Sec.~\ref{SM-supp:accel}), adds some overhead. 
Of note, one may cluster local environments at any degree of granularity (from discrete to fully continuous) using fuzzy classifiers such as {\tt DECAF}~\cite{DECAF}. Indeed, the latter readily identifies the Ni oxidation states discussed in the above conceptual \ce{Ni4O2} system, as well as in the stepped Ni surface, as discrete, distinct environments (SM, Sec.~\ref{SM-supp:decaf}).

In conclusion, we developed an adaptive DFTB approach that roots in the interpretation of the DFTB pseudo-atomic orbitals as static approximations of Kohn-Sham eigenstates. The latter suggests to locally tailor the employed confinement potentials to describe the varying orbital diffuseness in systems containing the same species in different chemical environments. In this work, we first provided a discrete proof-of-principle demonstration of this concept by assigning specifically optimized Slater-Koster parameters to atoms in different oxidation states. Although the presented applications are based on semi-local DFT references (for direct comparison with published parameters~\cite{mio-trans3d, Annies2021}), the adaptive construction itself is not tied to a particular reference level and can in principle be parameterized against higher-level functionals~\cite{Tammo2023}. 
The demonstrated smoothness of Slater-Koster parameters across oxidation environments paves the way to finer granularity, ultimately exploiting local atomic environment representations to predict SK parameters that continuously vary across compositions. 
We demonstrate this route by showing that a machine-learned site-resolved parameterization trained on a small dataset improves band structures over single-fit parameterizations across Ni--O binaries from Ni(0) to Ni(VI). The advance established here is the physical basis for local confinement adaptation across coexisting oxidation environments, rather than the learning machinery itself. Once this regularity is established, even a minimal regression can capture the environment-to-parameter mapping, shifting the challenge from architectural complexity to physical understanding. 

\begin{acknowledgements}
We gratefully acknowledge funding through grants no. 03HY130H (AEM-Direkt), no. 03SF0785C (MacGyver) and no. 03SF0798B (ASCEND) of the Bundesministerium für Forschung, Technologie und Raumfahrt (BMFTR), the DFG Cluster of Excellence EXC 2089 e-conversion 2.0 and the computational and data resources provided by the Max Planck Computing and Data Facility (MPCDF). We gratefully acknowledge valuable discussions with Dr. King Chun Lai, Dr. Andrew Wong, and Dr. Aditya Kumar.
\end{acknowledgements}

The authors declare no competing interests.

\textit{Data availability}—All the input files, scripts and relevant output files will be made available in a public repository upon publication.

\bibliographystyle{apsrev4-2}
\bibliography{references}

\appendix*
\section*{End Matter}

Appendix A:\applabel{A}{app:dftb3} Charge-dependent Hubbard parameters, as in DFTB3~\cite{DFTB3} (or a hypothetical adaptive $U$), offer an alternative route to improving SCC charges, energies and forces. They act, however, only on the charge response through $\gamma$ in $H^1_{\mu\nu}$, leaving $H^0_{\mu\nu}$ and $S_{\mu\nu}$, and hence the bonding interactions, unchanged. Adapting the confinement instead corrects the environment-dependent localization of the basis itself, and thus the dominant zeroth-order interaction, which higher-order SCC terms cannot modify. Moreover, the orbital (and density) compression through $V_{\mathrm{conf}}$ is the only empirical ingredient of $E_{\mathrm{el}}$, whereas $U$ is obtained from first principles and carries direct physical meaning. Environment dependence is therefore more naturally introduced through the confinement than through $U$.

Appendix B:\applabel{B}{app:gauss}\setcounter{equation}{0}\renewcommand{\theequation}{B\arabic{equation}} The band structure of the Ni$_4$O$_2$ trial system features multiple overlapping bands, making point-wise comparison between DFT and DFTB non-trivial. Conventional eigenvalue-based metrics such as the mean absolute error (MAE) can be ambiguous in such dense band manifolds containing crossings and near-degeneracies. We therefore introduce the Gaussian difference map, which compares broadened spectral intensity directly in $(k,E)$ space without requiring explicit band matching. The resulting Gaussian accuracy is sensitive to differences in the position and dispersion of spectral features near the Fermi level and is used here as a complement to the conventional MAE. The Gaussian difference map is constructed as follows. The DFT reference and DFTB band structures are each truncated to a chosen energy window around the Fermi level ($-2.5$ to $+5.0$~eV for the Ni$_4$O$_2$ analysis; $-5$ to $+5$~eV for the DOVE benchmark in Fig.~3. The wider window for the benchmark accommodates high-oxidation-state compounds whose principal near-Fermi features extend beyond $-2.5$~eV). Each band structure is then discretized on a uniform grid in $(k, E)$ space, represented as a scalar intensity field, and broadened with a 2D Gaussian kernel $\mathcal{G}_{\sigma_k, \sigma_E}$ with independent broadening widths $\sigma_k$ and $\sigma_E$ along the $k$-space and energy axes, respectively. Each grid point intensity is normalized by the maximum of the broadened intensity scale $B_\mathrm{max}$, rescaling all intensities to $[0, 1]$:

\begin{equation}
    \tilde{B}_i
    = \frac{\bigl(\mathcal{G}_{\sigma_k, \sigma_E} \ast B\bigr)_i}
           {B_\mathrm{max}} \,.
\end{equation}

The point-wise absolute difference between the normalized DFT and DFTB spectra yields the Gaussian difference map. A perfect band reproduction produces a uniformly vanishing map; residual discrepancies appear as regions of finite intensity (SM, Sec.~\ref{SM-supp:gaussmap}). From the map we define the Gaussian accuracy score $\mathcal{A}$ as
\begin{equation}
    \mathcal{A} = 1 - \mathcal{E} \,,
    \qquad
    \mathcal{E}
    = \frac{1}{N} \sum_{i=1}^{N}
      \bigl\lvert
        \tilde{B}_i^{\,\mathrm{DFTB}}
        - \tilde{B}_i^{\,\mathrm{DFT}}
      \bigr\rvert \,,
    \label{eq:accuracy_score}
\end{equation}
where $\mathcal{E}$ is the mean Gaussian error (MGE) averaged over all $N$ grid points in the map. The score is bounded $\mathcal{A} \in [0, 1]$, with $\mathcal{A} = 1$ corresponding to perfect reproduction. Results for the Ni$_4$O$_2$ system are reported in Table~\ref{table:scores}, alongside the conventional MAE for comparison. For completeness, a conventional MAE-based comparison across the Materials Project benchmark is reported in the SM (Fig.~\ref{SM-suppfig:mae_violin}), confirming that DOVE outperforms all other schemes also by this measure, while the Gaussian accuracy additionally evaluates the deviations in $(k,E)$ space.
\begin{figure}[t]
    \centering
    \includegraphics[width=0.95\linewidth]{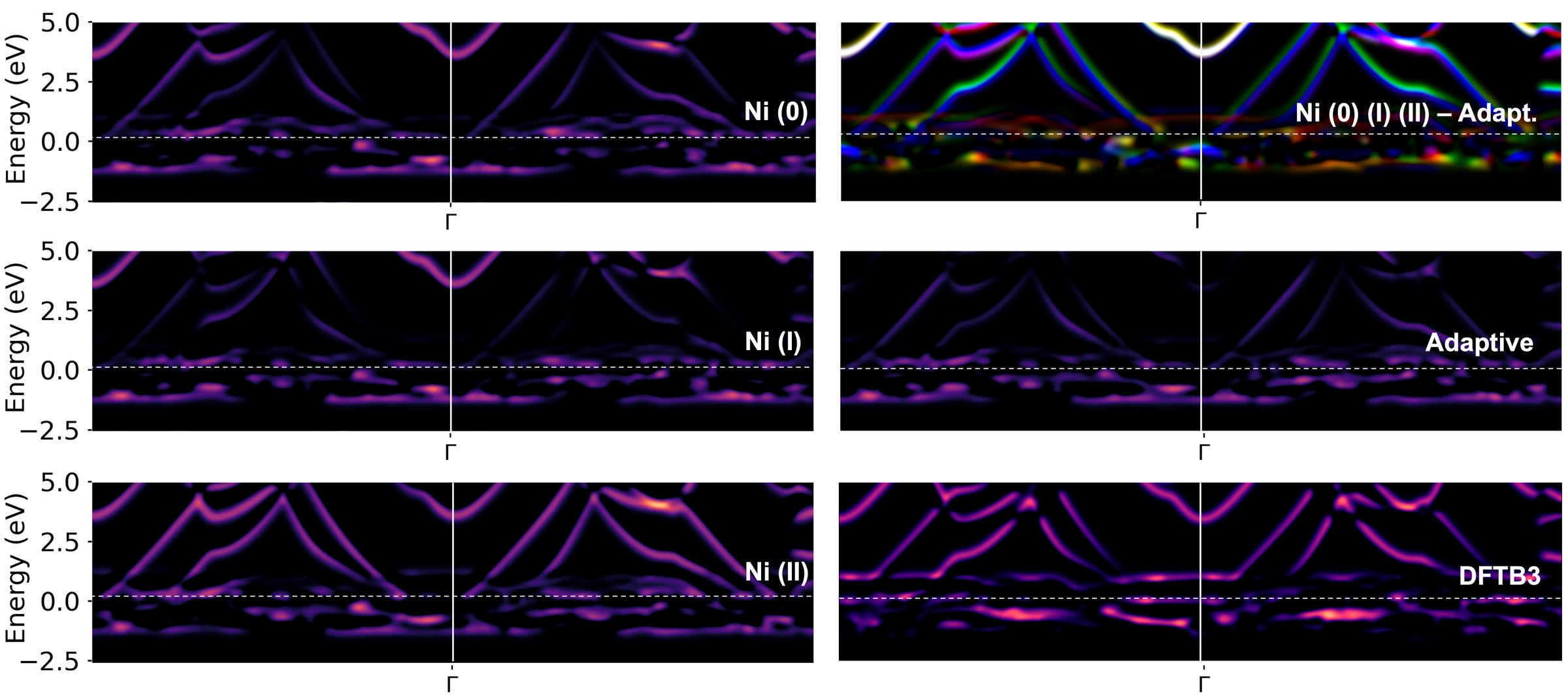}
    \caption{Gaussian difference maps for \ce{Ni4O2} calculated with different Ni DFTB parameters. The RGB figure represents the stacking of highlighted mismatch bands, which provides a joint picture of which features where the single-type parameterizations deviate from the adaptive calculation.(R, G, B: deviation of the Ni(0), Ni(I), and Ni(II) single-type parameterizations from the adaptive calculation, respectively.)}
    \label{fig:EM-3}
\end{figure}

\begin{table}[h!]
\centering
\caption{MGE and MAE for the \ce{Ni4O2} band structures under different parametrization schemes.}
\label{table:scores}
\begin{ruledtabular}
\begin{tabular}{cccccc}
                & Ni(0)& Ni(I)& Ni(II)& DFTB3     & Adaptive  \\
\hline
 MGE (\%)& 4.84& 6.03& 7.91& 7.32&3.76\\
 MAE (eV)& 0.120& 0.150& 0.162& 0.134&0.0972\\ 
\end{tabular}
\end{ruledtabular}

\end{table}

More granular information is provided by RGB analysis as shown in Fig.~\ref{fig:EM-3}. Here, Gaussian difference maps are constructed for each single type band structure with respect to the adaptive calculation and mapped onto red (Ni(0)), green (Ni(I)) and blue (Ni(II)) color channels. Areas where the single-type parameters deviate from the adaptive calculation are therefore highlighted in yellow, purple, cyan, or white, depending on which and how many single-type parameters deviate more from the adaptive approach for that particular feature. 

\begin{figure}[t]
    \centering
    \includegraphics[width=0.95\linewidth]{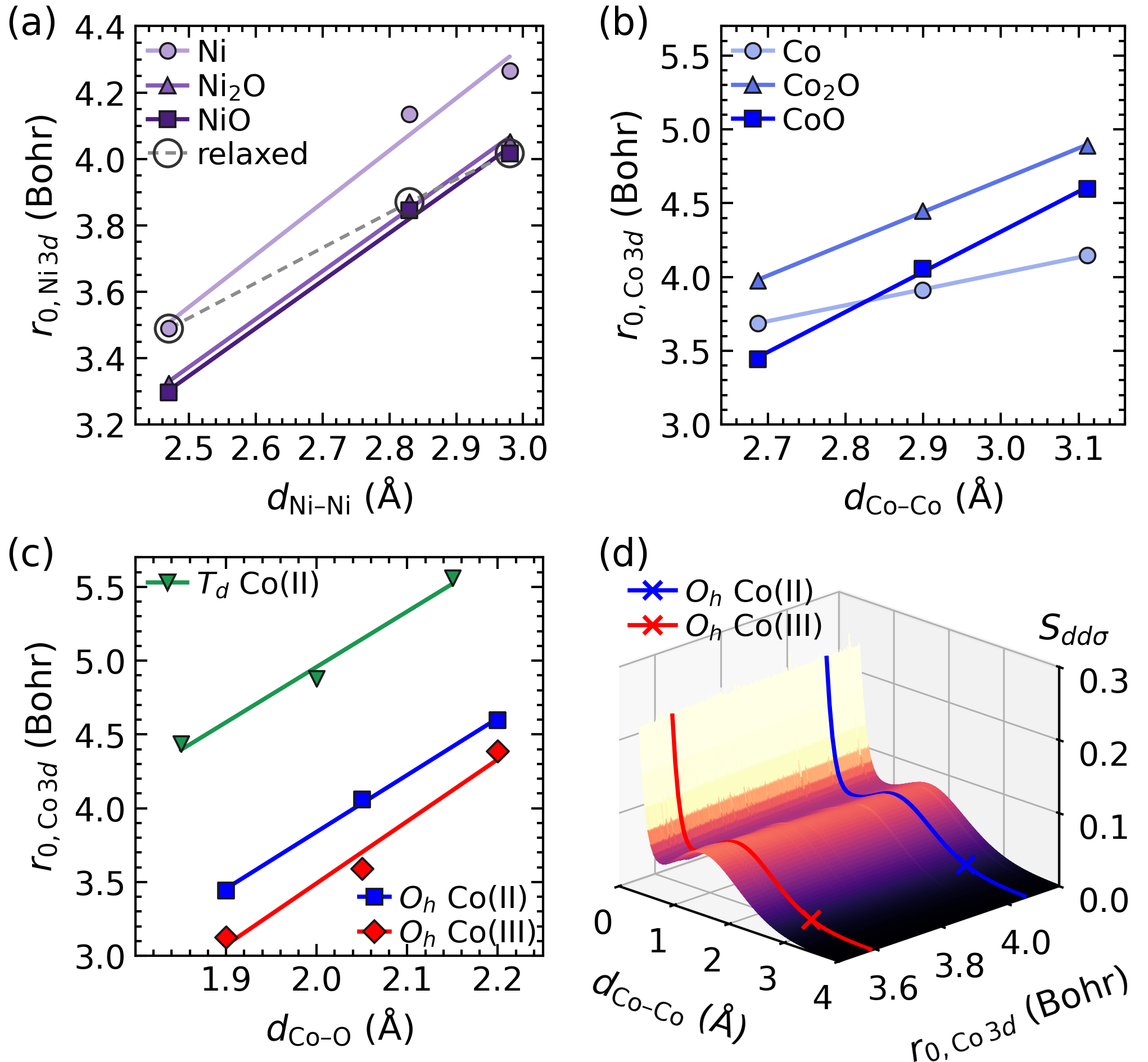}
    \caption{(a)~$r_{0,Ni_{3d}}$ for Ni, \ce{Ni2O}, and NiO versus the nearest-neighbor Ni--Ni distance; the three equilibrium systems are highlighted and connected with dashed lines. (b)~$r_{0,Co_{3d}}$ for Co, \ce{Co2O}, and CoO versus the nearest-neighbor Co--Co distance. (c)~$r_{0,Co_{3d}}$ versus $d_{\mathrm{Co-O}}$ for tetrahedral Co(II), octahedral Co(II), and octahedral Co(III). (d)~Co--Co $S_{dd-\sigma}$ for the parameter sets sampled in the octahedral Co(II) optimization, ordered by $r_{0,Co_{3d}}$. All lines in (a)--(c) are least-squares guides.}
    \label{fig:apx-2}
\end{figure}

Appendix C:\applabel{C}{app:linear} To elucidate the relationship between the optimal DFTB parameters and the varying local environment, here specifically the oxidation state, we address the apparent contradiction between orbital compression and diffuseness trends with oxidation noted in the main text. The key lies in bulk cell expansion. Despite the simplicity of our prototype \ce{NiO_x} systems, bulk volume—i.e., the equilibrium interaction distance—remains a crucial co-factor in orbital interactions. The nearest-neighbor Ni--Ni distance increases with oxidation, from 2.47~\AA{} (Ni) via 2.83~\AA{} (\ce{Ni2O}) to 2.98~\AA{} (NiO). Hence, we systematically investigate multiple trial systems within varying nearest-neighbor interacting distances (lattice parameters) based on the original Ni(0), Ni(I) and Ni(II) motifs. We apply particle swarm optimization to parameterize those trial systems, probing the coupling of DFTB parameters with spatial and chemical environments. As Fig.~\ref{fig:apx-2}a shows, $r_{0, Ni_{3d}}$ increases with $d$ (diffuseness) while nickel oxidation introduces a translational descent (compression).

In the simplest nickel oxide systems, spatial and chemical factors decouple, enabling an approximated linear expression for the confinement radius parameter in DFTB as \( r_0 = a d + b \), where \( d\) denotes the nearest-neighbor interacting distance, \( a \) captures the spatial dependence, and \( b \) encodes the chemical oxidation state as shown by the dashed line connecting the \{oxidation state, lattice constant\} points. For more complex environments, this local form is expected to generalize into non-linear regression.

The same analysis is extended to Co oxides with different oxidation states and coordination symmetry. For all investigated Co systems, the confinement exponents are kept fixed (SM, Sec.~\ref{SM-supp:coo_refs}). As shown in Fig.~\ref{fig:apx-2}(b), the Co, \ce{Co2O}, and CoO constructed from the same parent fcc framework give different responses to the spatial interacting distance $d_{Co-Co}$, where \ce{Co2O} and CoO respond similarly, while bulk Co follows a distinct trend. Figure~\ref{fig:apx-2}(c) further separates the effects of oxidation state and coordination symmetry by comparing $T_d$/\(O_h\) Co(II) and $O_h$ Co(II)/Co(III). Despite changing each factor independently, the three cases show similar spatial responses with slopes 3.75, 3.85, and 4.20~Bohr/\AA{}, respectively. 
Thus, neither oxidation state nor coordination symmetry alone determines the observed response. Instead, the similar behavior across the complete Co--O environments is consistent with the systematic response being associated with the Co--O bonding environment. Metallic Co shows a clearly different response with the same fixed confinement exponents, excluding the fixed exponent itself as the origin of this regularity. 
Figure~\ref{fig:apx-2}(d) shows the Co--Co $S_{dd-\sigma}$ integral for the parameter sets sampled in the octahedral Co(II) PSO, ordered by $r_{0,{\rm Co}_{3d}}$. As in the Ni case in Fig.~\ref{fig:SK_integral}a, the evolution is again smooth, suggesting that the smoothness is a general property of SK integrals rather than system-specific.

To test whether this regularity suffices to support site-resolved adaptive SK parameters beyond the prototypic Ni--O family, we consider mixed-valence spinel \ce{Co3O4}, whose primitive cell simultaneously contains two distinct coordination environments (tetrahedral Co(II) and octahedral Co(III)).
For each Co site and assigning method, the confinement parameter $r_0$ is obtained by interpolating a least-squares fit through the three reference optima of the corresponding motif in Fig.~\ref{fig:apx-2}c at the equilibrium distance, without knowing any \ce{Co3O4} electronic-structure information. Using these predicted parameters, we compare the DFTB band structures with the corresponding DFT result (SM, Sec.~\ref{SM-supp:co3o4}). The site-resolved adaptive scheme yields a band MAE of 0.241~eV over a large energy window from $-15$ to $+10$ eV, improving upon the uniform $T_d$- and $O_h$-derived assignments, which give MAE values of 0.387 and 0.276~eV, respectively.

Together, these results extend the physical foundation established for Ni--O to more diverse and complex oxides: the optimal confinement responds systematically to the local chemical and spatial environment, while the detailed form of this response depends on the bonding environment and cannot be represented by a single per-element linear relation. 

\end{document}


\title{Supplemental Material to\\ Adaptive Slater Koster Parameters:\\ Crossing Oxidation States with Density Functional Tight Binding}

\author{Yihua Song}
    \affiliation{Fritz Haber Institute of the Max Planck Society, Faradayweg 4-6, 14195 Berlin, Germany}

\author{Artem Samtsevych}
    \affiliation{Fritz Haber Institute of the Max Planck Society, Faradayweg 4-6, 14195 Berlin, Germany}

\author{Anton Beiersdorfer}
    \affiliation{Fritz Haber Institute of the Max Planck Society, Faradayweg 4-6, 14195 Berlin, Germany}

\author{Tobias Melson}
    \affiliation{Max Planck Computing and Data Facility, Gießenbachstraße 2, 85748 Garching b. München, Germany}

\author{Christoph Scheurer}
    \affiliation{Fritz Haber Institute of the Max Planck Society, Faradayweg 4-6, 14195 Berlin, Germany}
    
\author{Karsten Reuter}
    \affiliation{Fritz Haber Institute of the Max Planck Society, Faradayweg 4-6, 14195 Berlin, Germany}

\author{Chiara Panosetti}
    \email{panosetti@fhi.mpg.de}
    \affiliation{Fritz Haber Institute of the Max Planck Society, Faradayweg 4-6, 14195 Berlin, Germany}

\date{\today}

\maketitle

\section{\label{sec:Computational Detail}Computational Details}

All the input files, scripts and relevant output files will be made available in a public repository upon publication.

\subsection{Density Functional Theory Calculations}
The DFT reference band structures were obtained via the Atomic Simulation Environment ({\tt ASE}) \cite{HjorthLarsen2017-ase} with the Vienna \textit{ab initio} simulation package ({\tt VASP})~\cite{kresse1996} using a projector augmented plane wave (PAW)~\cite{PAW-PhysRevB.50.17953} basis set within the Perdew–Burke–Ernzerhof (PBE)~\cite{PBE.PhysRevLett.77.3865} functional. The plane wave cutoff threshold was set to 520 eV. Self consistent cycles were converged to a total energy threshold of $10^{-5}$ eV and applying Gaussian smearing with a width of 0.1~eV. All calculations employed a well-converged k-point density of 0.05~\AA{}$^{-1}$. To simplify the parametrization process and for consistency with previously published parameters~\cite{mio-trans3d}, the reference band structures used for the confinement optimization were calculated without spin polarization nor DFT+U corrections. Additional spin-polarized references used only for the transferability tests are described in Sec.~\ref{supp:spin}. The geometries for Ni and NiO were obtained from the Materials Project \cite{material_project_Jain2013}, while the geometry for \ce{Ni2O} was constructed starting from a $(2 \times 1 \times 1)$ NiO supercell with one oxygen atom removed, then the atomic coordinates and cell parameters locally relaxed using the Broyden–Fletcher–Goldfarb–Shanno (BFGS) \cite{HjorthLarsen2017-ase} optimizer to a force convergence of 0.05 eV/\AA{} while keeping the cell shape fixed. Tersoff–Hamann STM simulations were performed using {\tt vaspkit} \cite{vaspkit}.

\subsection{Density Functional Tight Binding Calculations}
All the DFTB calculations were performed with the {\tt dftb+} package version 24.1, compiled with MPI support. K-point density and band paths for all calculations were set as the same as the corresponding DFT reference calculations. Shell resolved SCC calculations were converged to a threshold of \mbox{1e-5 eV} applying a 0.0001 Hartree electronic temperature (corresponding to ca. 31.7 K). For the spin-polarized calculations, the colinear type of initial spin is employed and relaxed using the spin constants obtained from {\tt SKProgs}\cite{Tammo2023}. Tersoff–Hamann STM simulations were performed using {\tt waveplot} offered with {\tt dftb+}.

\subsection{DFTB parameterization}
All parameterizations were performed by DFTB Slater-Koster Optimizer ({\tt DSKO})~\cite{dsko}, with Particle Swarm Optimization (PSO)~\cite{PSO488968} as global optimizer, {\tt SKProgs}~\cite{Tammo2023} as Slater-Koster file (\texttt{.skf}) generator, and {\tt dftb+}~\cite{Hourahine2020} as calculator. The wave function compressions were optimized shell resolved using a power confinement for Ni, Co, O and a Woods-Saxon confinement for Li, C. All the other DFTB parameters (eigenvalues, $U$ and density compression) were kept fixed at the default values provided by {\tt DSKO}. The two-center integrals were calculated using density superposition for Ni, O and potential superposition for Li, C. PSO was run over 80 iterations of 144 particles for all systems, except metallic Li and graphitic C (50 iterations of 96 particles). The repulsive parameterization for lithium-intercalated graphite is performed by {\tt GPrep} \cite{Chiara2023}. 

\subsection{Nudged Elastic Band Calculations}
Nudged Elastic Band (NEB) \cite{JNSSON1998-NEB-original,neb-climbing} calculations were performed using the default implementation in {\tt ASE} \cite{HjorthLarsen2017-ase} with {\tt dftb+} as a calculator. The initial and final states were locally relaxed to a force convergence threshold of 0.05 eV/\AA{} using the BFGS optimizer. Bands of 10 and 17 images, for the armchair and zigzag terminations respectively, were interpolated using the image dependent pair potential (idpp). Constraints were applied to selected atoms in the intercalated bulk region to avoid numerical instabilities in the optimization. A first NEB optimization was performed without climbing images to 0.5 eV/\AA{} with BFGS as the optimizer. Then, a stricter NEB optimization with climbing images to 0.05 eV/\AA{} convergence was performed with FIRE \cite{Bitzek2006-fire} as the optimizer. The converged NEB images were recalculated single-point for all other schemes mentioned in the main text.

\section{\label{sec:NiOx}\ce{NiO_x}}

\subsection{\ce{NiO_x} DFTB parameterization results}

The detailed results of the parameterizations are reported in \Cref{supptab:dsko}. The optimal values for the dominant parameter $r_{0, Ni_{3d}}$ are highlighted in bold. These increase steadily with the oxidation state, while the non-dominant parameters show less significant or negligible trends. 

\begin{table}[h!]
\centering
\caption{Optimal DFTB parameters for Ni in different oxidation states. The optimal values for the dominant parameter $r_{0, Ni_{3d}}$ are highlighted in bold.}
\label{supptab:dsko}
\begin{ruledtabular}
\begin{tabular}{ccccccc}
System  & \textbf{$r_{0, Ni_{3d}}$}    &\textbf{$\sigma_{Ni_{3d}}$}& $r_{0, Ni_{4s}}$   &\textbf{$\sigma_{Ni_{4s}}$}& $r_{0,Ni_{4p}}$  &\textbf{$\sigma_{Ni_{4p}}$}\\
\hline
Ni&\textbf{3.49} &7.37& 2.97 &6.56&3.42 &7.82\\ 
$\ce{Ni2O}$&\textbf{3.87} &4.90& 2.98 &4.63&3.56 &10.93\\  
NiO&\textbf{4.01} &7.99& 3.49 &4.44&3.35 &6.48\\
\end{tabular}
\end{ruledtabular}

\end{table}

\begin{figure}[h]
    \centering
    \includegraphics[width=0.9\linewidth]{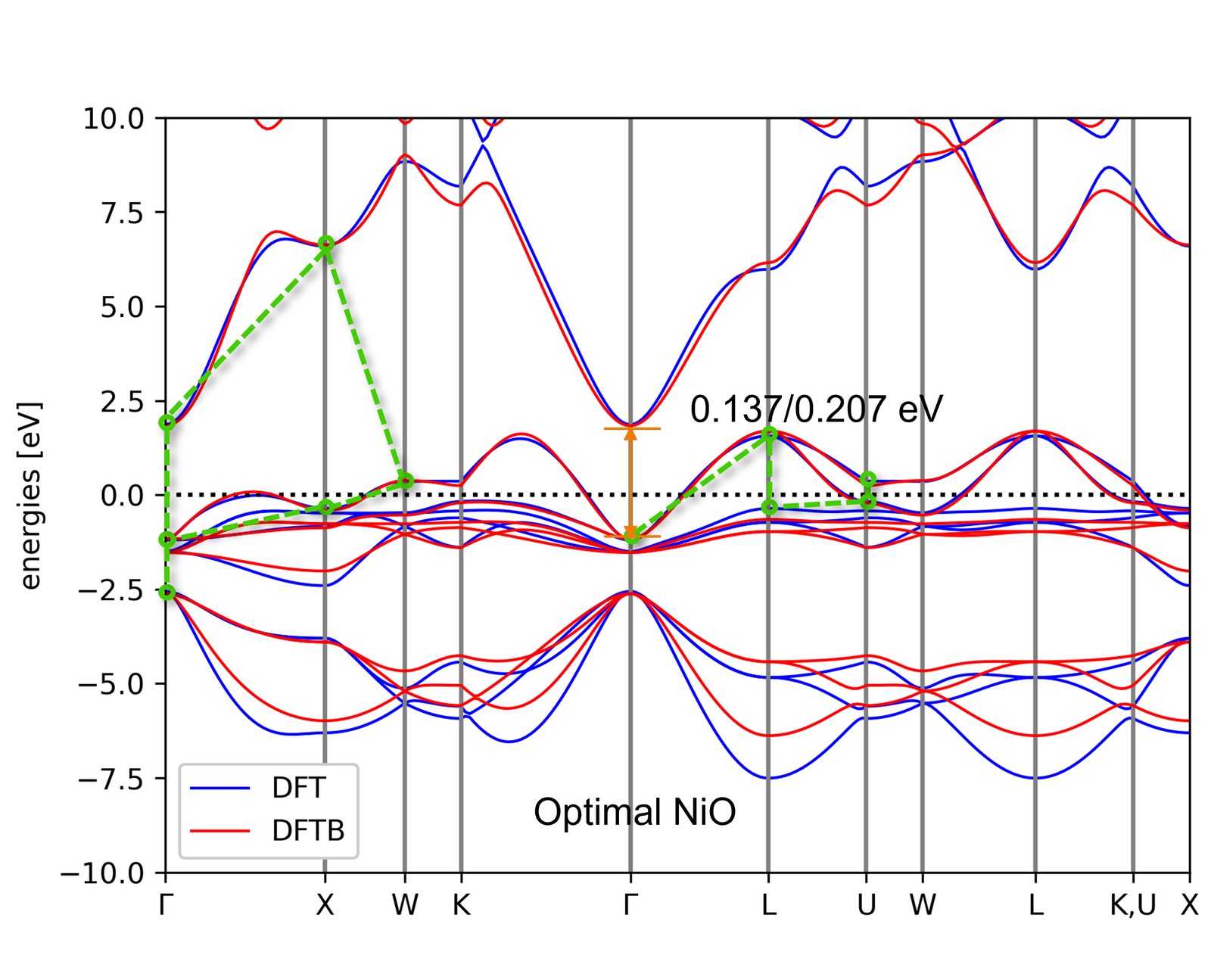}
    \caption{Band structure comparisons of the NiO bulk system within our optimal Ni(II) parameters.}
    \label{suppfig:NiO-band-gap}
\end{figure}

Fig. \ref{suppfig:NiO-band-gap} shows the reference DFT-PBE band structure (blue) and the optimal DFTB band structure (red) for \ce{NiO} as well as the employed decoding mask (green dashed lines), designed to target especially the region around the Fermi level with little focus on the lower bands. 

\subsection{\ce{NiO} band gap}

\begin{figure}[h]
    \centering
    \includegraphics[width=0.9\linewidth]{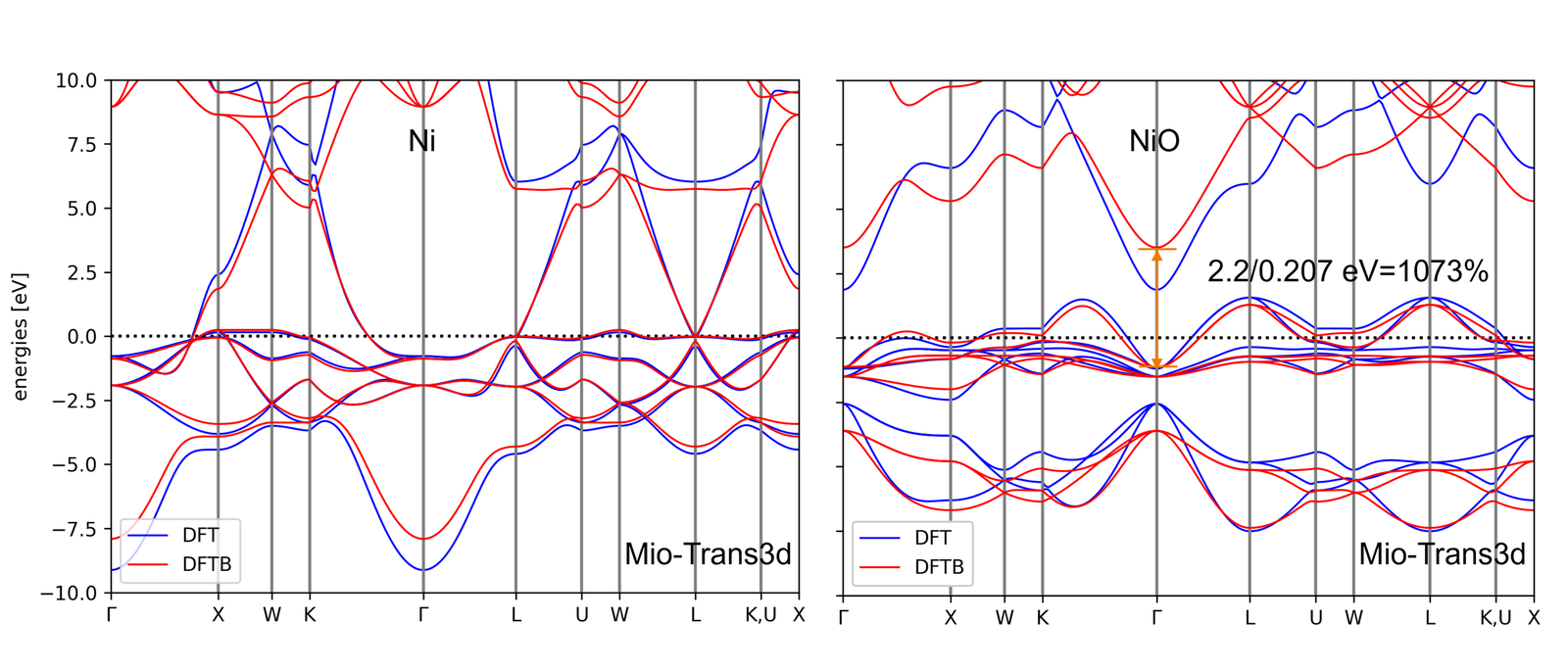}
    \caption{Band structure comparisons of Ni and NiO with Mio-Trans3d parameters, showing a poor description of the NiO band gap.}
    \label{suppfig:mio3d}
\end{figure}

As Fig.~\ref{suppfig:mio3d} shows, trans-3d Ni parameters~\cite{mio-trans3d}, optimized on the bulk Ni reference, reproduce the metallic Ni band structure correctly but fail for NiO: the predicted band gap of 2.221\,eV overestimates the PBE value of 0.207\,eV by about an order of magnitude. Our parameters optimized for NiO instead yield 0.138\,eV ({\em cf.} Fig.~\ref{suppfig:NiO-band-gap}), close to the PBE reference (which, in passing, underestimates the experimental gap, as expected for a semi-local functional). Parameters optimized for Ni can thus not be used for NiO and vice versa, underscoring the need for locally adapted electronic parameters.

The transferability of the optimized parameters was assessed by calculating band structures for expanded and compressed geometries. Band agreements remain reasonable for variations of the lattice parameters up to ca. 10\%, beyond which the deviations become too large. Transferability to spin-polarized calculations is discussed in Sec.~\ref{supp:spin}.

\subsection{Optimal DFTB Parameters for \ce{NiO_x}}

\begin{figure}
\centering
\includegraphics[width=0.9\linewidth]{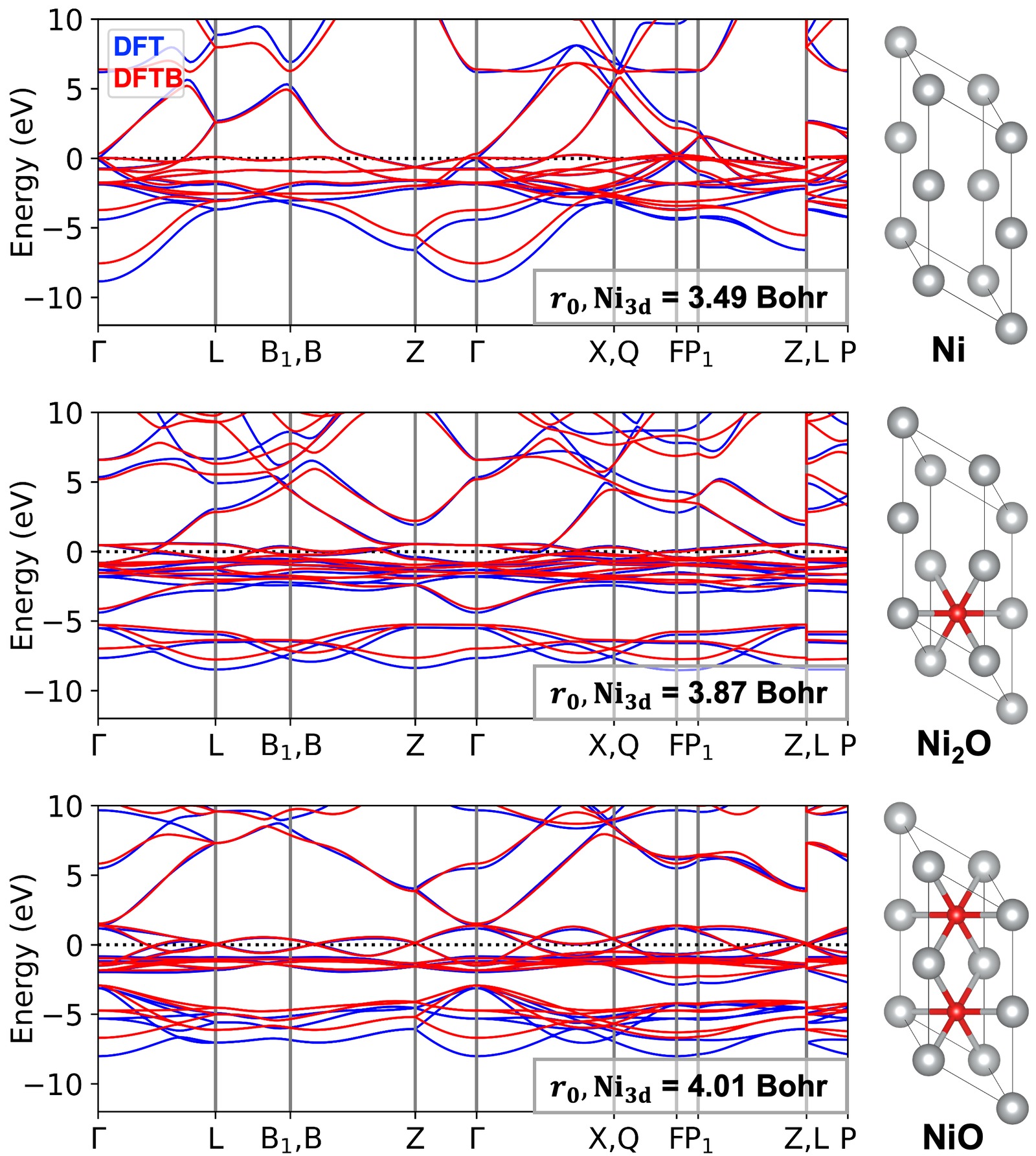}
  \caption{Band structure comparison between the DFT reference and the DFTB calculation with optimized parameters for each oxidation state. The corresponding optimal onset radius $r_{0, {\rm Ni}_{3d}}$ for Ni \textit{3d} is shown for bulk Ni, \ce{Ni2O}, and NiO, corresponding to the formal oxidation states Ni(0), Ni(I), and Ni(II), respectively. All band structures are calculated in the same cell shape, corresponding to the $(2 \times 1 \times 1)$ supercell of Ni as illustrated to the right, so to be shown piecewise along the same band path.}
    \label{suppfig:SI-params of NiOx}
\end{figure}
Fig.~\ref{suppfig:SI-params of NiOx} compares the DFT reference and DFTB band structures obtained with the parameters optimized for each oxidation state.

\subsection{Transferability of the confinement parameters to spin-polarized calculations}
\label{supp:spin}

\begin{figure}[h]
\centering
\includegraphics[width=\columnwidth]{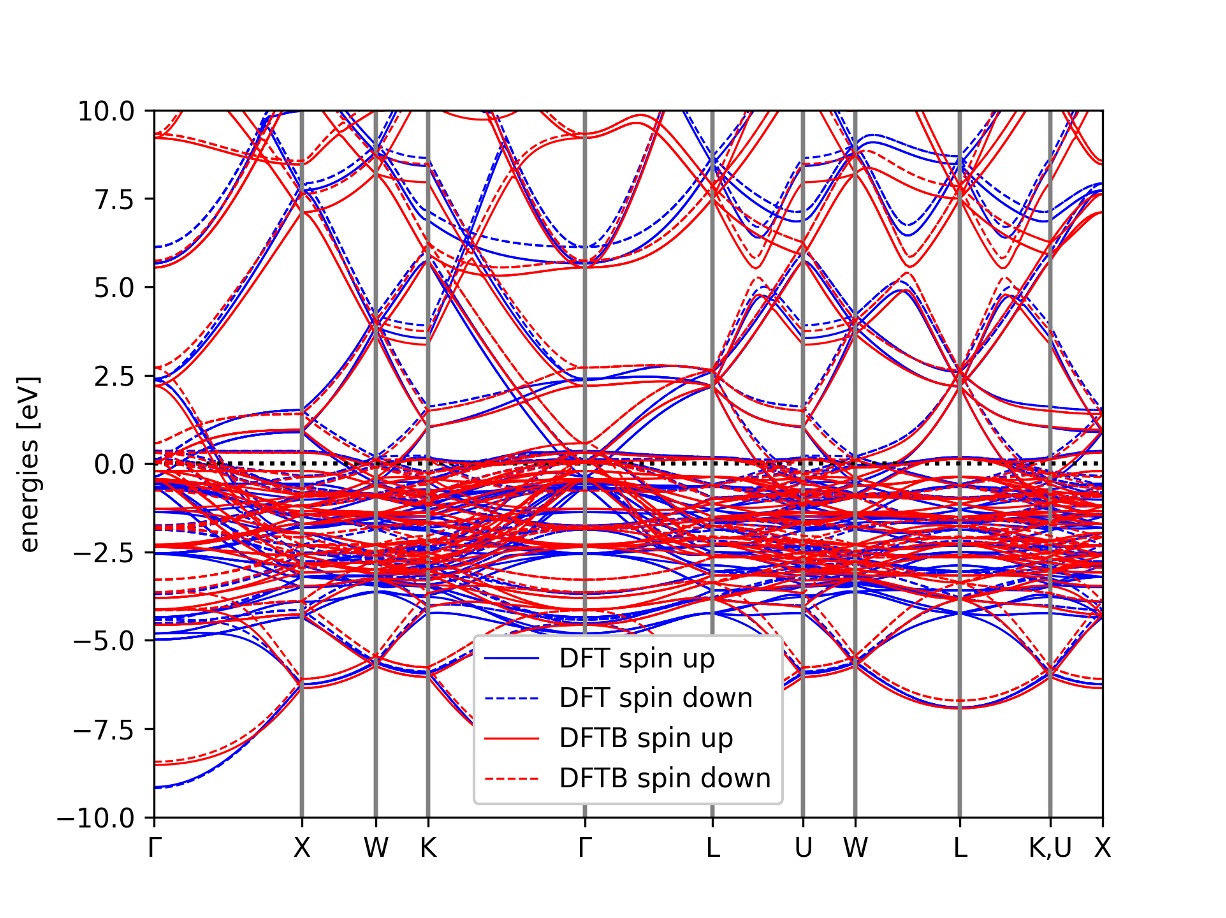}
\caption{Spin-resolved band structure of ferromagnetic Ni (4-atom 
cell) comparing spin-polarized DFT-PBE (blue) and DFTB (red) using 
Ni(0) parameters optimized against the non-spin-polarized 
reference. Solid and dashed lines denote majority (spin-up) and 
minority (spin-down) channels, respectively.}
\label{suppfig:spin_pol}
\end{figure}

To assess the portability of confinement parameters optimized against non-spin-polarized references, we first consider ferromagnetic Ni. We perform collinear spin-polarized DFTB calculations for ferromagnetic Ni (4-atom cell) using the optimal Ni(0) parameters obtained from the non-spin-polarized parametrization. Figure~\ref{suppfig:spin_pol} compares the resulting spin-resolved band structure against spin-polarized DFT-PBE. Despite the absence of spin polarization in the optimization target, the DFTB bands reproduce the spin-split DFT reference well, particularly in the occupied $d$-band manifold below the Fermi level. The exchange splitting between majority and minority channels is qualitatively captured, indicating that the confinement parameters encode orbital localization behavior that carries over to the spin-polarized case.

We next consider NiO in the AFM-II order, with ferromagnetic (111) Ni planes stacked antiparallel. Calculations use the four-atom rhombohedral cell at the rocksalt lattice constant $a = 4.217$~\AA{}. The complete Ni/O confinement parameter set obtained from the original non-spin-polarized NiO parameterization is retained without refitting. Spin-polarized DFTB2 calculations with the fully localized limit (FLL) on-site correction~\cite{Hourahine2007} are compared with two spin-polarized references: PBE (case~1) and PBE$+U$ in the Dudarev formulation with $U = 6.5$~eV on Ni~$3d$ (case~2).

In case~1, the O~$2s$ and $2p$ corrections are taken from a fit to ferromagnetic NiO and used without fitting to the AFM-II reference. In case~2, the FLL parameters are fitted to the AFM-II PBE$+U$ band structure, while the confinement parameters remain fixed. The corrections are listed in Table~\ref{supptab:NiO_spin}; no correction is applied to Ni in case~1, and FLL corrections are applied to Ni $3d$, $4s$, $4p$ in case~2. The fitted Ni~$3d$ value, $U{-}J = 0.311$~Ha (8.46~eV), is similar in magnitude to the reference $U = 6.5$~eV, although the DFTB and PAW projections differ. These FLL parameters are distinct from the SCC Hubbard parameters of the underlying DFTB2 parameterization.

\begin{table}[h]
\caption{AFM-II NiO calculations with the original non-spin-polarized Ni/O confinement parameters retained unchanged. The listed $U{-}J$ values are FLL corrections in DFTB; dashes indicate that no FLL correction is applied to the corresponding shell. Gaps are evaluated along the sampled band path. Ni moments are local magnitudes obtained from VASP site projections and DFTB Mulliken analysis, respectively.}
\label{supptab:NiO_spin}
\begin{ruledtabular}
\begin{tabular}{lcc}
 & Case 1 & Case 2 \\
\hline
DFT reference & PBE & PBE$+U$ \\
DFT $U$, Ni~$3d$ (eV) & -- & 6.5 \\
DFTB $U{-}J$, Ni~$3d$ (Ha) & -- & 0.311 \\
 DFTB $U{-}J$, Ni~$4s$ (Ha)& -- &0.011\\
 DFTB $U{-}J$, Ni~$4p$ (Ha)& -- &0.019\\
DFTB $U{-}J$, O~$2s$ (Ha) & 0.547 & 0.254 \\
DFTB $U{-}J$, O~$2p$ (Ha) & 0.321 & 0.197 \\
Gap, DFT / DFTB (eV) & 1.138 / [0.925] & 3.236 / [3.308] \\
Ni moment, DFT / DFTB ($\mu_\mathrm{B}$) & 1.36 / [1.31] & 1.77 / [1.81] \\
\end{tabular}
\end{ruledtabular}
\end{table}

\begin{figure}[h]
\centering
\includegraphics[width=\columnwidth]{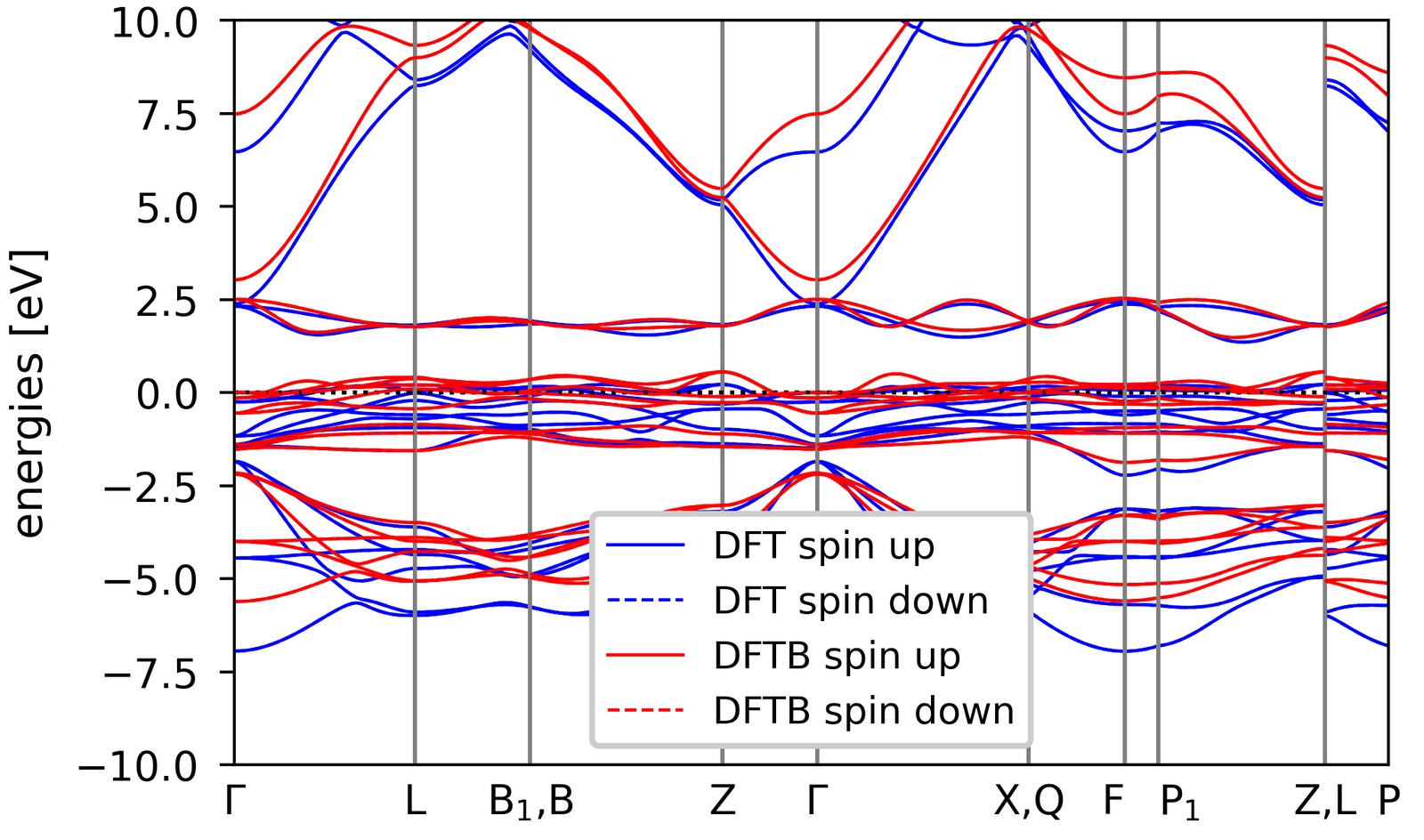}
\caption{Spin-resolved band structure of AFM-II NiO comparing PBE (blue) and DFTB2 with FLL corrections on O (red), using the retained non-spin-polarized Ni/O confinement parameters (case~1 in Table~\ref{supptab:NiO_spin}). Solid and dashed lines denote spin-up and spin-down channels, which coincide in AFM-II. Energies are relative to each method's highest occupied state at $\Gamma$.}
\label{suppfig:NiO_spin_PBE}
\end{figure}

\begin{figure}[h]
\centering
\includegraphics[width=\columnwidth]{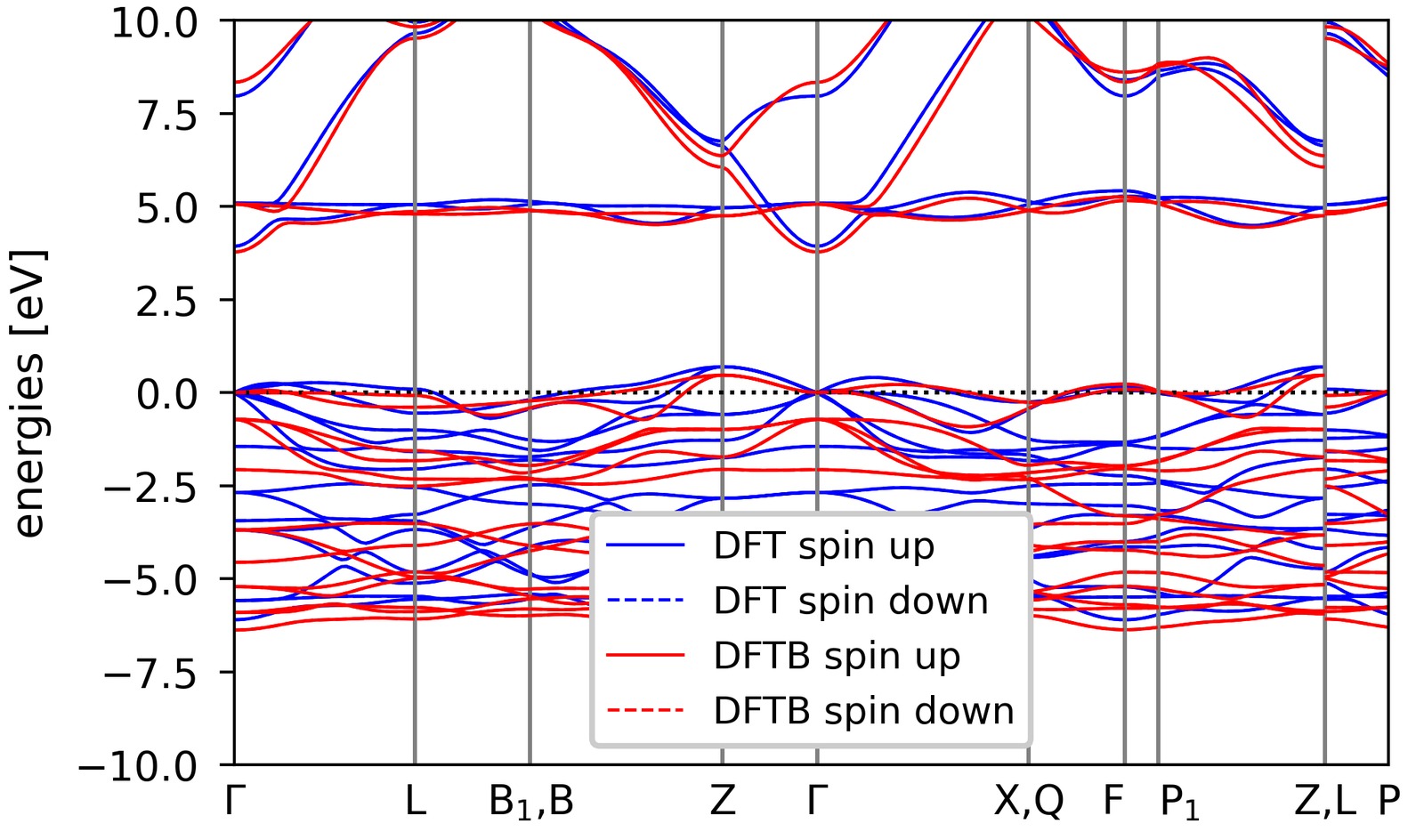}
\caption{Spin-resolved band structure of AFM-II NiO comparing PBE$+U$ ($U = 6.5$~eV on Ni~$3d$; blue) and DFTB2 with fitted FLL corrections on Ni~$3d$, $4s$, $4p$, O~$2s$, and O~$2p$ (red), using the retained non-spin-polarized Ni/O confinement parameters (case~2 in Table~\ref{supptab:NiO_spin}). Solid and dashed lines denote spin-up and spin-down channels, which coincide in AFM-II. Energies are relative to each method's highest occupied state at $\Gamma$.}
\label{suppfig:NiO_spin_PBEU}
\end{figure}

Figures~\ref{suppfig:NiO_spin_PBE} and \ref{suppfig:NiO_spin_PBEU} compare the spin-resolved band structures. Both calculations converge to insulating AFM-II solutions with equal and opposite Ni moments, and the gaps and local moments are close to the corresponding DFT values (Table~\ref{supptab:NiO_spin}). Deviations remain in the valence-band dispersion, particularly in case~2. These results show that the confinement parameters obtained without spin remain usable when spin polarization and on-site corrections are introduced; the additional FLL parameters, rather than the confinement basis, are adjusted to the reference treatment.

\subsection{\ce{Ni4O2} band structure}
\label{supp:nio_bs}

\begin{figure}[h]
    \centering
    \includegraphics[width=0.9\linewidth]{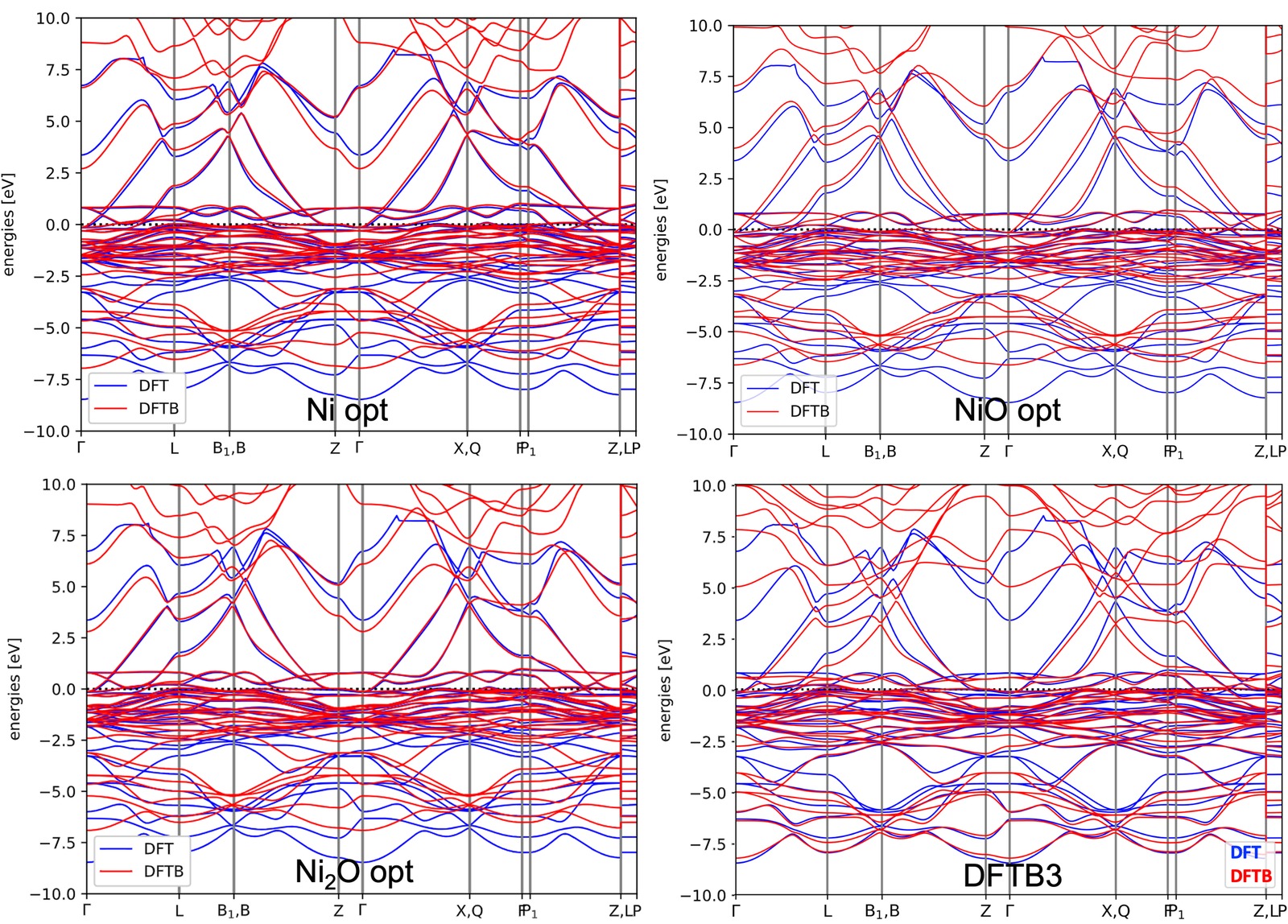}

    \caption{Band structure comparisons of the prototypic \ce{Ni4O2} system within the optimal Ni(x) (in DFTB2) and published Mio-Trans3d parameters (in DFTB3, with Hubbard derivatives from \cite{Nickel-DFTB3}). }
    \label{suppfig:DFTB3-Ni4O2}
\end{figure}

Fig. \ref{suppfig:DFTB3-Ni4O2} shows the band structure comparison using different parametrizations, respectively, the optimal Ni, optimal \ce{Ni2O}, optimal NiO and trans-3d with DFTB3 corrections. As reported in the main text and further visualized with the Gaussian difference maps, none of the single-type parameterizations, including trans-3d with DFTB3 corrections, reproduces the band structure equally well as the adaptive approach. Of note, trans-3d with DFTB3 shows a better agreement in the lower bands; however, those were not targeted at all in our PSO optimizations.

\begin{figure}[t]
    \centering
    \includegraphics[width=0.9\linewidth]{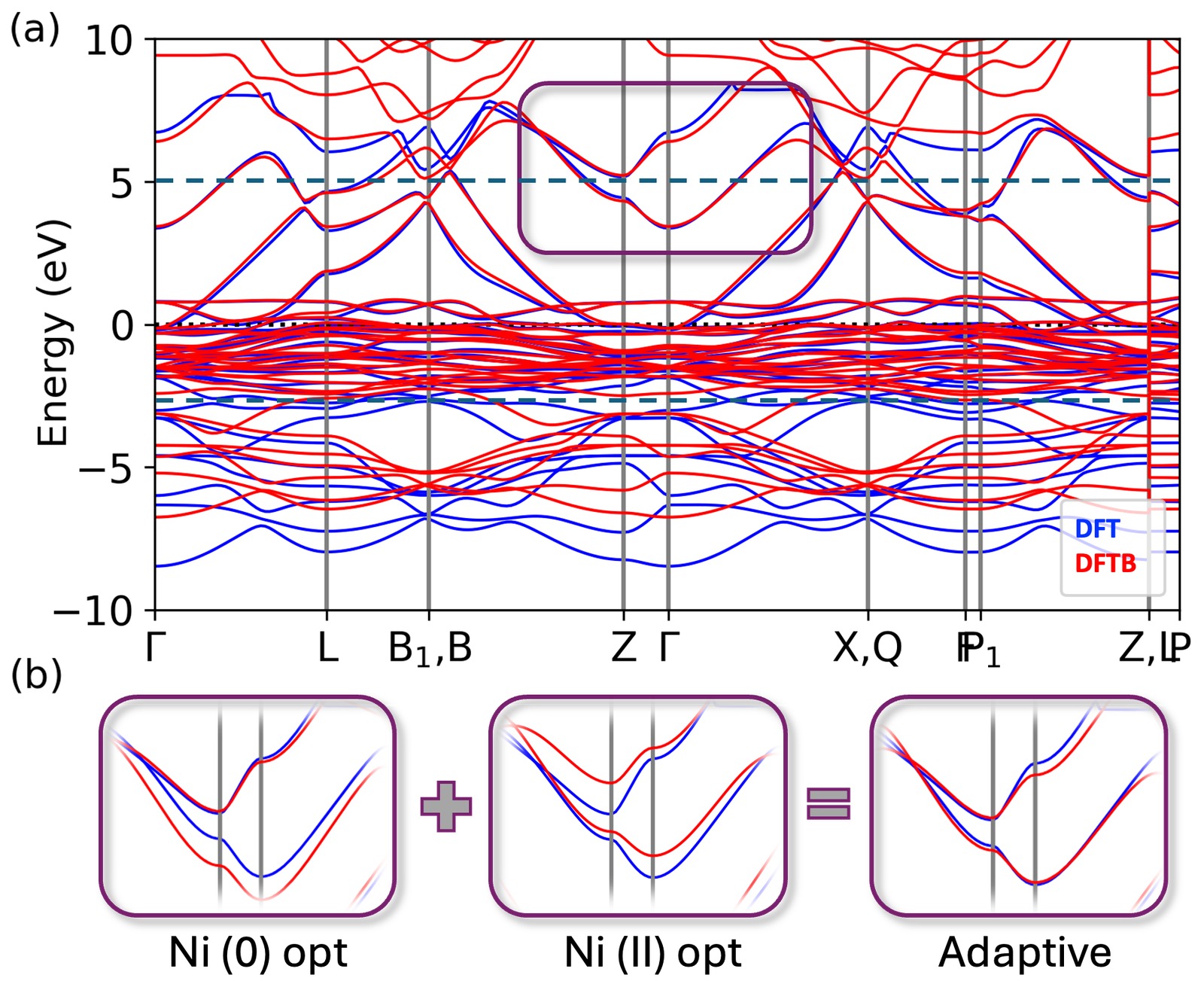}
    \caption{(a) \ce{Ni4O2} band structure comparison between DFT and DFTB with the adaptive approach.  Green dashed lines indicate the energy range at which the band structure is truncated to construct the Gaussian difference maps. (b) Schematic representation of a simplified mechanism of improvement of the electronic properties in the adaptive approach. For the isolated band, the single-type confinement for Ni(0) and Ni(II) produces too large and too small dispersion, respectively, while the adaptive parameterization produces the correct dispersion.
}
    \label{suppfig:closer-look}
\end{figure}

A closer inspection of the band structure suggests a possible simplified mechanism by which the local parameterization improves the description of electronic properties. In Fig.~\ref{suppfig:closer-look} (b), the adaptive band prediction is compared with the band structures obtained from "single-type" parameters, which shows that the optimal Ni(0) and the optimal Ni(II) parameters produce slightly stretched and slightly compressed bands respectively in the highlighted feature, while the adaptive parameterization produces the correct dispersion. Ni(0)-optimal parameters correspond to a globally less diffuse orbital, but yield a larger overlap at the relevant bond distance, while Ni(II)-optimal parameters correspond to a globally more diffuse orbital, but yield a smaller overlap at the relevant bond distance. As a result, consistently with "textbook" tight binding, a larger overlap corresponds to more dispersed bands, and {\em vice versa}. 

\subsection{PDOS and DBC Analysis}
\begin{figure}[h]
    \centering
    \includegraphics[width=0.95\linewidth]{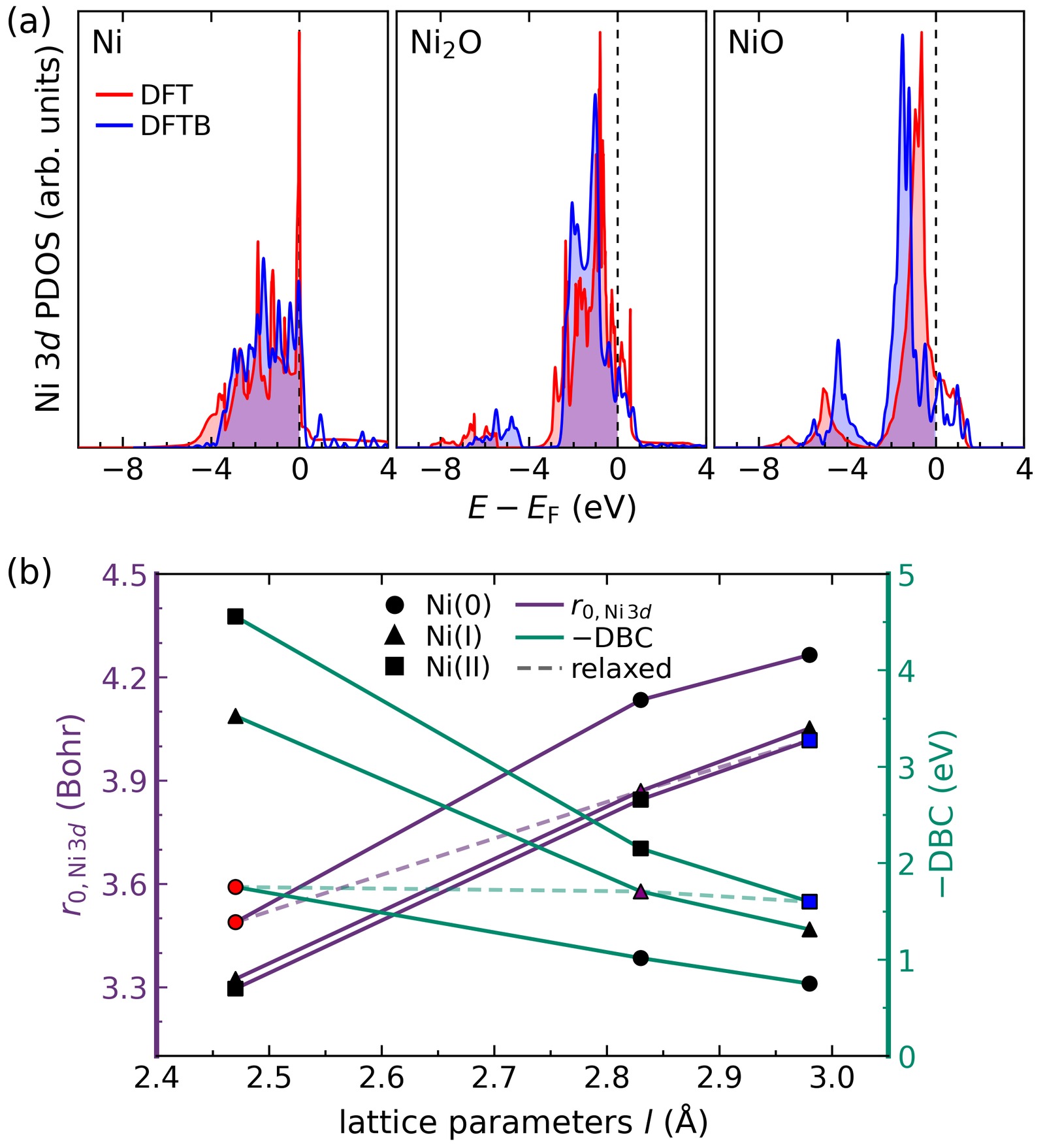}
    \caption{(a)~Non-spin-polarized Ni-\textit{3d} PDOS from DFT (red) and DFTB (blue) for Ni, \ce{Ni2O}, and NiO. (b)~Optimized $r_{0,{\rm Ni}_{3d}}$ (purple, left axis) and $-$DBC (green, right axis) for Ni(0) (circles), Ni(I) (triangles), and Ni(II) (squares) at $l$=2.47, 2.83, and 2.98~\AA{}; colored markers connected by dashed guides indicate the relaxed Ni, \ce{Ni2O}, and NiO structures.}
    \label{suppfig:DBC}
\end{figure}

We further employ PDOS and \textit{d}-band center (DBC) analysis~\cite{Norskov200071} to track the evolution of electronic properties across the \ce{NiO_x} ($x=0,0.5,1$) systems. PDOS calculations show a good fit with DFT calculations within our optimal parameters, as expected. As shown in Fig.~\ref{suppfig:DBC}a, the comparison of Ni-\textit{3d} PDOS in the three \ce{NiO_x} systems shows a more compressed \textit{d} band in the oxidized NiO bulk.

The DBC is defined as the first moment of the DFT Ni-\textit{3d} PDOS $\rho_d(E)$,
\[
    \varepsilon_d =
    \frac{\int E\,\rho_d(E)\,dE}
         {\int \rho_d(E)\,dE}.
\]
The DBC of \ce{NiO_x} bulk systems shows an increasing trend with the oxidation process, as shown by the relaxed sequence in Fig.~\ref{suppfig:DBC}b. The key lies in bulk cell expansion. Despite the simplicity of our prototype \ce{NiO_x} systems, bulk volume---i.e., the equilibrium interaction distance---remains a crucial co-factor in orbital interactions. The lattice parameter $l$ (the nearest-neighbor Ni--Ni distance in this construction) increases with oxidation, from 2.47~\AA{} (Ni) via 2.83~\AA{} (\ce{Ni2O}) to 2.98~\AA{} (NiO).

Hence, we systematically investigate multiple trial systems within varying lattice parameters based on the original Ni(0), Ni(I) and Ni(II) motifs, and calculate their PDOS and DBC with DFT for uni-variate analysis. Table~\ref{tab:S3} reports the corresponding DBC values for each oxidation state and each $l$. The green lines in Fig.~\ref{suppfig:DBC}b show that the DBC levels decrease with oxidation and increase with the enlarging cell. Since Fig.~\ref{suppfig:DBC}b plots $-$DBC, these trends appear with reversed sign: the green lines rise with oxidation and fall with increasing $l$. The relaxed sequence combines both trends.

We apply particle swarm optimization to parameterize those trial systems, probing the coupling of DFTB parameters with spatial and chemical environments. Table~S4 reports the corresponding parameterization results. As Fig.~\ref{suppfig:DBC}b shows, $r_{0,{\rm Ni}_{3d}}$ (purple lines) increases with $l$ (diffuseness) while nickel oxidation introduces a translational descent (compression). The detailed numbers are shown in Table~\ref{tab:S4}. The relaxed systems combine both effects, yielding the spatial and chemical trend discussed in End Matter Appendix~\ref{main-app:linear}.

\begin{table}[h!]
\centering
\caption{\textit{d}-band center (DBC) analysis for Ni($x$) oxide systems at varying nearest-neighbor Ni--Ni distances. $l_n$ denotes the relaxed distance for oxidation state $n$: 2.47, 2.83, and 2.98~\AA{} for $l_0$, $l_{\rm I}$, and $l_{\rm II}$, respectively.}
\label{tab:S3}
\begin{ruledtabular}
\begin{tabular}{cccc} 
 DBC(eV)& $l_0$& $l_I$&$l_{II}$\\
\hline
Ni(0)& \textbf{-1.754}& -1.017&-0.752 \\ 
Ni(I)& -3.527& \textbf{-1.708}&-1.314\\ 
Ni(II)&-4.559& -2.153&\textbf{-1.601} \\
\end{tabular}
\end{ruledtabular}

\end{table}

\begin{table}[h!]
\centering
\caption{Parametrization results of the dominant $r_{0,{\rm Ni}_{3d}}$ for Ni($x$) oxide systems at the same $l_n$ as in Table~S3.}
\label{tab:S4}
\begin{ruledtabular}\begin{tabular}{cccc} 
\textbf{$r_{0, Ni_{3d}}/Bohr$}& $l_0$& $l_I$&$l_{II}$\\
\hline
Ni(0)& \textbf{3.489}& 4.134&4.265\\ 
Ni(I)& 3.324& \textbf{3.870}&4.052\\ 
Ni(II)&3.296& 3.845&\textbf{4.017}\\
\end{tabular}
\end{ruledtabular}

\end{table}

\subsection{\label{supp:decaf}Identification of Ni types with fuzzy classification (DECAF)}

\begin{figure}[h]
\centering
\includegraphics[width=0.9\linewidth]{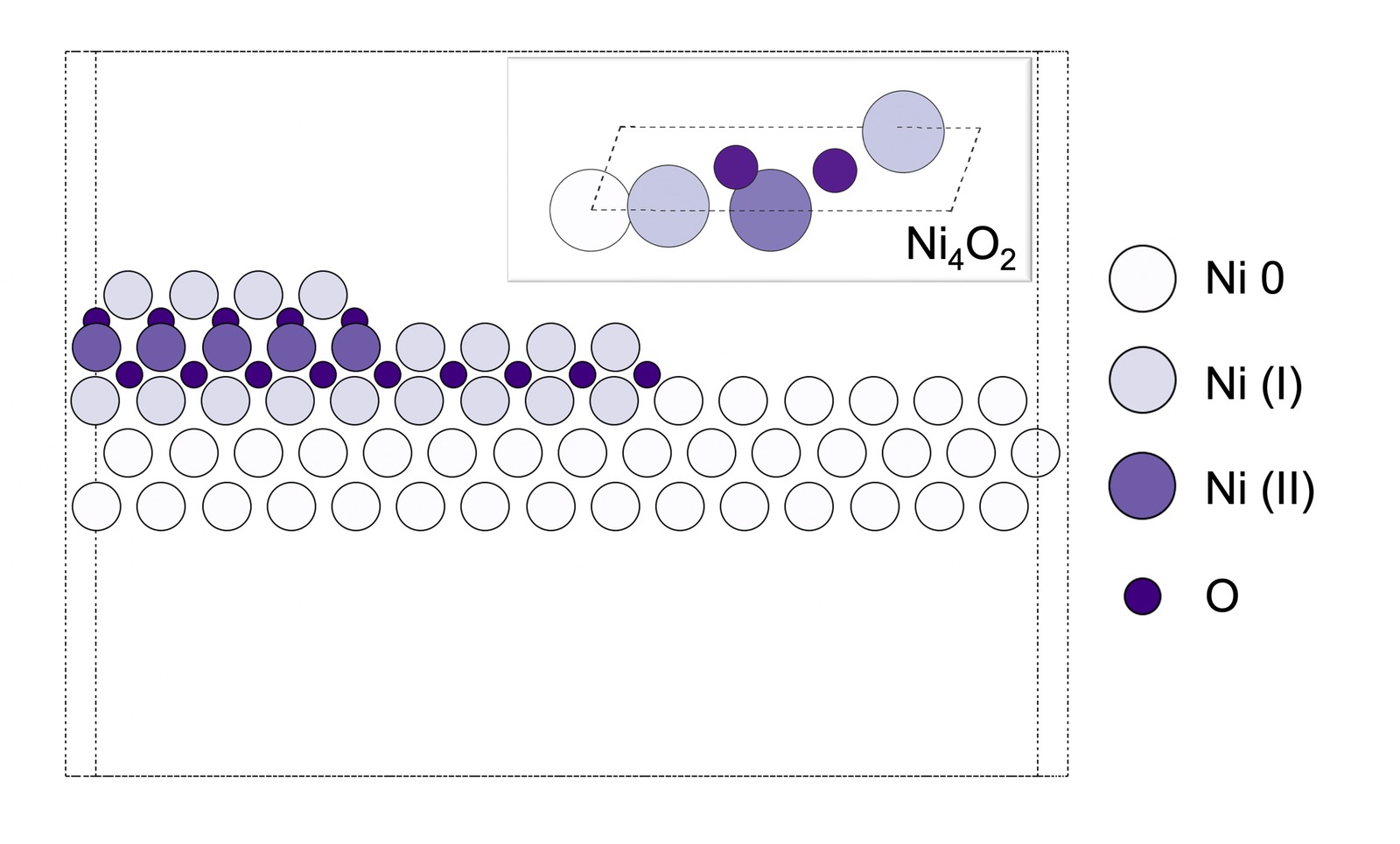}
\caption{Side view of the NiO stepped surface overlayed with the view of the prototypic \ce{Ni4O2}, with colored atoms indicating the assignment of discrete types of Ni($x$) in the adaptive DFTB scheme. The classification of Ni($x$) types are performed by {\tt DECAF}.}
\label{suppfig:stepped-surface}
\end{figure}

Fig.~\ref{suppfig:stepped-surface} illustrates the application of the adaptive DFTB approach to the prototypic \ce{Ni4O2} system and the NiO stepped-surface oxide within the STM simulation. Given the observed smoothness of the SK parameters, we employ {\tt DECAF}\cite{DECAF} as an ML-driven classifier to assess the robustness and potential for broader ML integration within the adaptive DFTB framework. {\tt DECAF} provides a fuzzy classification scheme for identifying equivalent atoms in complex materials and molecules. Using Haar--Wavelet bandwidth optimization, the three discrete Ni($x$) types are correctly classified at the optimal bandwidth in the "all-types-in-one" \ce{Ni4O2} system, and their stability is confirmed across a range of bandwidths in the subsequent scan test. We therefore applied this framework to assign the Ni($x$) types in the stepped‑surface system. As shown in Fig.~\ref{suppfig:stepped-surface}, the nickel atoms are successfully clustered according to their local environments in agreement with chemical intuition.

\subsection{\label{supp:mulliken}Mulliken population analysis}

As an independent consistency check on the Gaussian difference metric, we compare Ni Mulliken populations between DFT and DFTB across the benchmark set, following the same workflow as in the main-text population comparison. On the DFT side, populations are obtained by projecting the converged VASP plane-wave wavefunctions onto contracted Slater-type orbitals using LOBSTER~\cite{Ertural2019,Nelson2020}, with charge spilling below 2\% in all cases. On the DFTB side, Mulliken populations arise natively from the localized pseudo-atomic basis and enter the self-consistent charge cycle directly.

Since the LOBSTER projection employs extended Slater-type orbitals while DFTB uses confined pseudo-atomic orbitals, orbital-resolved populations are subject to systematic differences that reflect the projection basis rather than the electronic structure~\cite{Mulliken1955}. This effect is most pronounced for the diffuse Ni~$4p$ channel, which serves primarily as a polarization function in the DFTB minimal basis. We therefore restrict the comparison to the Ni~$4s$ and $3d$ populations, which carry the dominant valence character and are directly governed by the confinement potential optimization central to this work.

As shown in Fig.~\ref{main-fig:SK_integral}b of the main text, DOVE (RMSE\,=\,0.113\,e) reproduces the DFT reference populations significantly more accurately than both published DFTB3 parameters (0.264\,e) and the best optimized single-type \ce{Ni2O} parameterization (0.170\,e). The reported RMSE is evaluated over the contributing Ni atoms by pairing the DFT and DFTB valence populations for the same atom. We note that absolute Mulliken populations are inherently basis-set dependent and thus not directly comparable between the two representations; however, the relative ranking of the methods is robust, as the DFT reference is obtained from an identical projection for all comparisons. The largest deviation occurs for the Ni(0) in \ce{Ni4O}, where all valence electrons reside in $s$ and $d$ channels in the LOBSTER projection, while the DFTB basis redistributes a fraction into $4p$, amplifying the projection mismatch at vanishing charge transfer.

\subsection{\label{supp:gaussmap}Gaussian difference map}

The broadening width $\sigma_E$ is chosen to resolve the relevant spectral features near the Fermi level. For the Ni$_4$O$_2$ analysis ($-2.5$ to $+5.0$~eV window), $\sigma_E = 40$ ($\approx 0.16$~eV) is used on a $7200 \times 2400$ grid ($k \times E$). For the Materials Project benchmark ($-5$ to $+5$~eV window), the broadening is increased to $\sigma_E = 
80$ ($\approx 0.33$~eV) to accommodate the wider spread of band gaps across the high-oxidation-state compounds in the library. In both cases, the Gaussian accuracy ranking among methods is preserved under moderate variations of $\sigma_E$. Before constructing the maps, each saved DFT and DFTB spectrum is shifted by its own Fermi energy.

\subsection{Overlap matrix conditioning}
\label{sec:overlap_condition}

The generalized eigenvalue problem $\mathbf{H}\mathbf{c} = 
\varepsilon\,\mathbf{S}\mathbf{c}$ requires the overlap matrix 
$\mathbf{S}$ to be positive definite and well-conditioned. Its condition number $\kappa(\mathbf{S}) = \lambda_\mathrm{max} / 
\lambda_\mathrm{min}$ bounds the sensitivity of the computed eigenvalues to perturbations in $\mathbf{S}$: a relative perturbation $\|\delta\mathbf{S}\|/\|\mathbf{S}\|$ can shift the 
eigenvalues by up to $\kappa(\mathbf{S})$ times that amount. A large $\kappa$ therefore signals that small numerical errors—or, in the present context, the mixing of basis functions from inequivalent confinement potentials—could be amplified into unreliable eigenvalues.

In the adaptive approach, atoms in different oxidation states are assigned
pseudo-atomic orbitals obtained from different confinement potentials. Since
these orbitals enter the same generalized eigenvalue problem (eq.~(2) in the
main text), a natural concern is whether the resulting overlap matrix
$\mathbf{S}(\mathbf{k})$ becomes ill-conditioned when basis functions from
different confinements are mixed. We address this by computing the condition
number $\kappa(\mathbf{S}) = \lambda_\mathrm{max}/\lambda_\mathrm{min}$ at
each $\mathbf{k}$-point for the converged overlap matrix, comparing the
adaptive and single-type calculations as well as published DFTB3
parameters\cite{mio-trans3d,Nickel-DFTB3}.

\begin{figure}[h]
\centering
\includegraphics[width=\columnwidth]{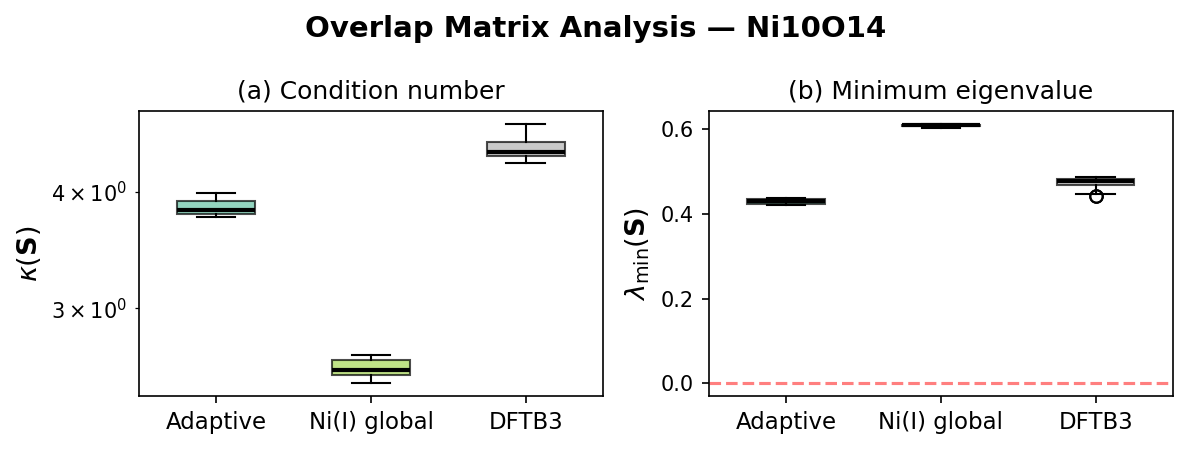}
\caption{Distribution of the overlap matrix condition number
$\kappa(\mathbf{S})$ (left) and minimum eigenvalue
$\lambda_\mathrm{min}(\mathbf{S})$ (right) across all $\mathbf{k}$-points for
Ni$_{10}$O$_{14}$, comparing the adaptive approach, globally applied Ni(I)
parameters, and published DFTB3 parameters. All methods yield single-digit
condition numbers, with $\lambda_\mathrm{min}$ well above zero throughout.}
\label{suppfig:overlap_condition}
\end{figure}

Table~\ref{tab:overlap_condition} and Fig.~\ref{suppfig:overlap_condition} summarize the results for the prototypic Ni$_4$O$_2$
system and a representative Ni$_{10}$O$_{14}$ structure predicted by DOVE from the Materials
Project library. In all cases, $\kappa(\mathbf{S})$ remains in the single
digits and $\lambda_\mathrm{min}$ stays well above zero, confirming that the
overlap matrix is well-conditioned throughout the Brillouin zone. Notably, for
both cases, the adaptive parameterization yields lower condition numbers
than the published DFTB3 scheme. This demonstrates that mixing orbitals from different confinements does not introduce numerical instability in the generalized eigenvalue problem.

\begin{table}[h]
\centering
\caption{Overlap matrix condition numbers
$\kappa(\mathbf{S}) = \lambda_\mathrm{max}/\lambda_\mathrm{min}$ for
different parameterization schemes. $\bar{\kappa}$ denotes the mean over all
$\mathbf{k}$-points and $\lambda_\mathrm{min}^\mathrm{global}$ the smallest
eigenvalue encountered across the Brillouin zone.}
\begin{tabular}{l l c c c c}
\hline\hline
Structure & Method & $\kappa_\mathrm{min}$ & $\bar{\kappa}$ & $\kappa_\mathrm{max}$ & $\lambda_\mathrm{min}^\mathrm{global}$ \\
\hline
Ni$_4$O$_2$       & Adaptive     & 4.6 & 6.5 & 8.2 & 0.286 \\
                   & Ni(I) global & 2.8 & 3.6 & 4.3 & 0.435 \\
                   & DFTB3 & 5.3 & 7.3 & 8.4 & 0.293 \\
\hline
Ni$_{10}$O$_{14}$ & Adaptive     & 3.8 & 3.9 & 4.0 & 0.419 \\
                   & Ni(I) global & 2.5 & 2.6 & 2.7 & 0.602 \\
                   & DFTB3 & 4.3 & 4.5 & 4.7 & 0.441 \\
\hline\hline
\end{tabular}
\label{tab:overlap_condition}
\end{table}

\section{\label{supp:li}Lithium-intercalated graphite}

\subsection{\ce{Li_xC6} DFTB parameterization results}

Table \ref{supptab:opt_li} reports the optimal dominant parameters as optimized by {\tt DSKO} for metallic and intercalated lithium, as well as the adopted values for interfacial lithium as the average between the former two. The other parameters in the Woods-Saxon confinement potential model, i.e. the height \(W\) and the slope \(a\) are kept fixed across different Li types to the values optimized for metallic Li. 

\begin{table}[h!]
\centering
\caption{Optimal DFTB parameters for different Li types. Here, the parameters of interfacial Li are the mean values from the metallic and the intercalated Li.}
\label{supptab:opt_li}
\begin{ruledtabular}
\begin{tabular}{ccc}
Type & \textbf{$r_{0, Li_{2s}}$} & $r_{0, Li_{2p}}$   \\
\hline
Metallic&\textbf{6.102}& 6.821 \\ 
Interfacial&\textbf{5.361}& 6.807 \\ 
Intercalated&\textbf{4.620}& 6.794 \\
Published DFTB2 Li& \multicolumn{2}{c}{\textbf{5.77}} \\
\end{tabular}
\end{ruledtabular}

\end{table}

The parameters of interfacial Li are approximated by the arithmetic mean of the metallic and intercalated values, reflecting the coordination environment of the interfacial atom, intermediate between the two bulk endpoints. This is an explicit approximation within the discrete-type picture, not a prediction of the DOVE model.

\begin{figure}
    \centering
    \includegraphics[width=0.9\linewidth]{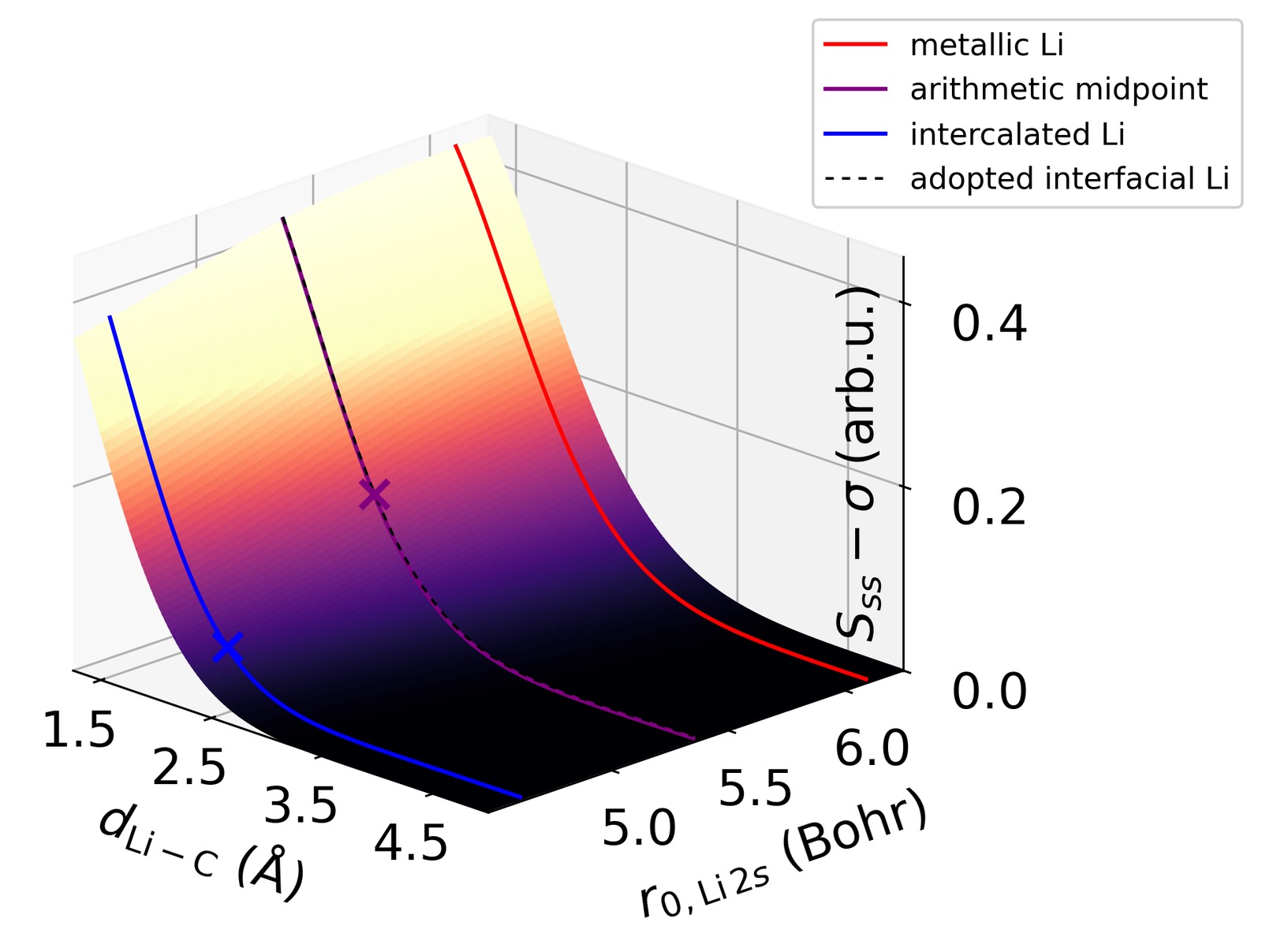}
    \caption{Li--C $S_{ss-\sigma}$ overlap integral as a function of the Li--C distance and the Li $2s$ confinement radius. Curves corresponding to the metallic-Li and intercalated-Li parameterizations are shown together with the arithmetic midpoint of the two confinement radii and the adopted interfacial-Li parameterization. The smooth variation of the integral across this interval supports the midpoint interpolation used for interfacial Li.}
    \label{suppfig:Li_midpoint}
\end{figure}

To check that this interpolation does not introduce an abrupt change in the electronic coupling, we evaluate the Li--C $S_{ss-\sigma}$ integral as a function of the Li--C distance and the Li $2s$ confinement radius (Fig.~\ref{suppfig:Li_midpoint}). The integral varies smoothly between the metallic- and intercalated-Li parameterizations, and the adopted interfacial parameterization closely follows the arithmetic-midpoint curve over the relevant Li--C distances. This provides a direct consistency check for using the midpoint confinement at the interface.

\subsection{Li insertion into graphite}

\begin{figure}[h]
    \centering
    \includegraphics[width=0.9\linewidth]{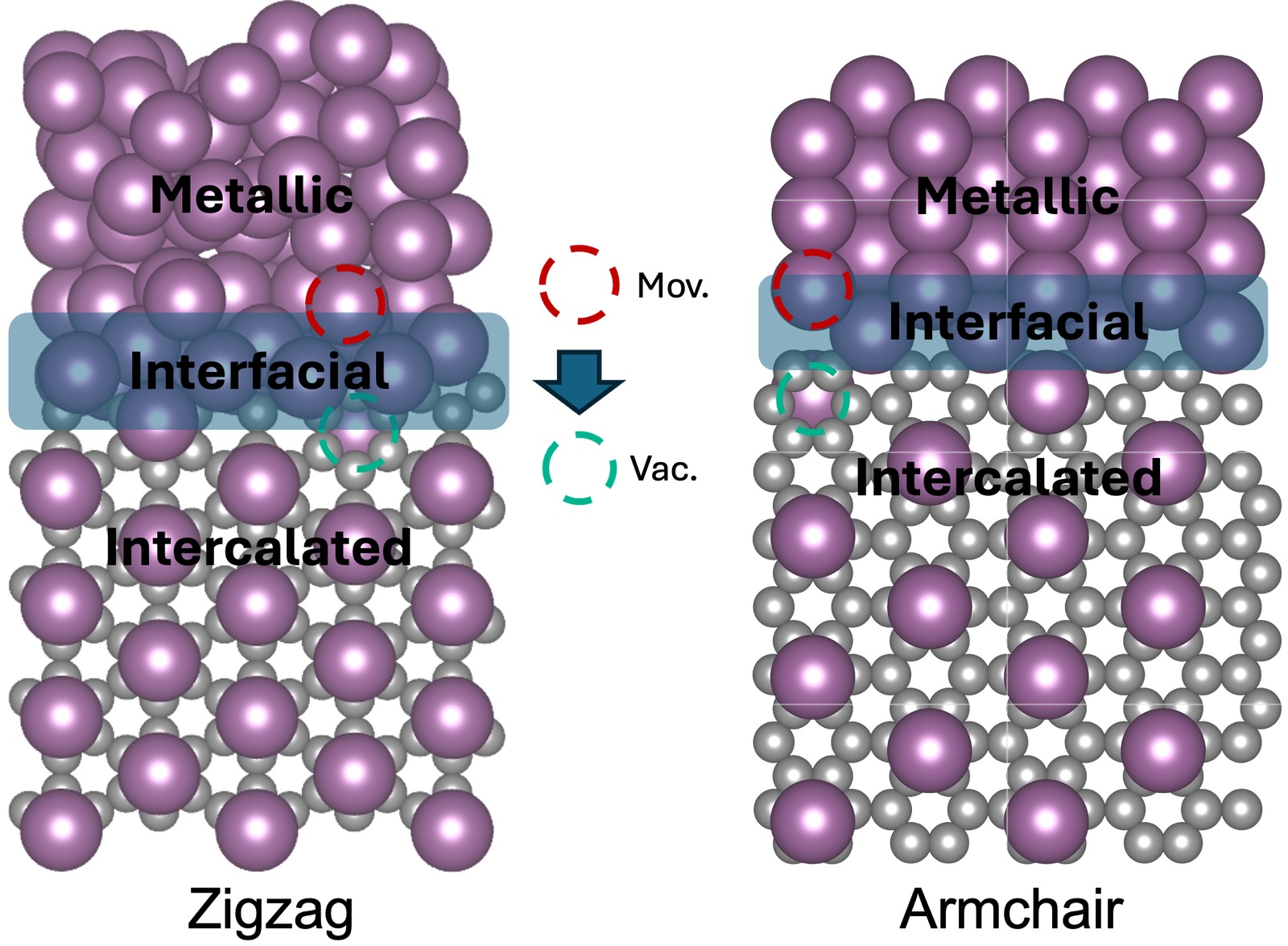}
    \caption{Geometry of lithium insertion paths across zigzag and armchair terminations of graphite. Lithium atoms are identified as metallic, the interfacial, and intercalated types. The red dashed circle marks the moving Li, while the green dashed circle marks the vacancy for insertion.}
    \label{suppfig:zigzag-geo}
\end{figure}

Fig. \ref{suppfig:zigzag-geo} summarizes the initial and final state geometries for the zigzag (left) and armchair (right) terminations, with one single Li atom moving from the metallic Li region (red circle) to a vacancy in the intercalation region (green circle). The moving atoms in both cases are selected by the lowest energy of the final state among all possible Li atoms. The figure also highlights the three different Li types, identified by both geometric and partial charge features.

The electronic-only adaptation retains the published DFTB2 repulsion, whereas combined adaptation employs three Li types of environment-dependent repulsive terms fitted with {\tt GPrep}\cite{Chiara2023}, corresponding to the discrete electronic adaption. Small discrepancies and discontinuities are expected to be solved by generalizing the approach to the continuum of chemical environments beyond discrete atomic types.


The resulting energy profiles are shown in Fig.~\ref{main-fig:applications}b of the main text. Since the geometries along the paths are obtained with published DFTB2 parameters, the endpoints are not necessarily minima at the DFT or adaptive DFTB level.
%
%

\section{DOVE Machine Learning Framework}

\subsection{\label{supp:training}Training set construction}

The training set for the DOVE Gaussian Process Regression (GPR) model
consists of 60 Ni--O binary structures containing no more than 4 atoms each.
These 60 structures were selected from a larger candidate pool generated from a genetic algorithm {\tt EZGA}~\cite{Juan2026}, by a systematic filtering procedure designed to sample the diversity of local atomic environments in the Ni--O binary configuration space while minimizing redundancy. The procedure combines fuzzy clustering of physical descriptors~\cite{DECAF}, used as a chemical environment discriminator, with an information-theoretic evaluation~\cite{Godden2000}. The fuzziness of the environment discriminator is adjusted to group sufficiently similar atomic environments, with one representative selected from each group. The informative entropy of the resulting subset is then evaluated. Further details of the selection methodology will be reported in an upcoming publication.
For each training structure, optimal DFTB confinement parameters were
obtained via {\tt DSKO}~\cite{dsko} using Particle Swarm Optimization
(PSO), yielding per-atom optimized values for the onset radius $r_0$ and decay
exponent $\sigma$ of the power-law confinement potential for each atomic
shell: Ni~$3d$, $4s$, $4p$ and O~$2s$, $2p$.

\subsection{Atomic environment descriptors}

Local atomic environments are represented using a modified Smooth Overlap of Atomic Positions (SOAP)~\cite{SOAP_PhysRevB.87.184115} descriptor as similarly used in DECAF\cite{DECAF}, computed independently for
Ni and O atoms. The SOAP hyperparameters (cutoff radius, number of radial
and angular basis functions, and Gaussian smearing width) were chosen to
ensure sufficient resolution of the local coordination environment while
maintaining computational efficiency. Separate descriptor matrices are
constructed for Ni atoms ($\mathbf{X}_{\mathrm{Ni}}$) and O atoms
($\mathbf{X}_{\mathrm{O}}$), reflecting the different chemical roles and
numbers of atoms per element across the training set.

\subsection{Gaussian Process Regression}\label{supp:gpr}

DOVE employs independent single-output GPR models for the onset radius $r_0$ of each atomic shell, yielding five models in total. Each GPR model uses a constant signal variance prefactor multiplied by a Mat\'ern kernel with smoothness parameter
$\nu = 3/2$:
\begin{equation}
k(\mathbf{x}, \mathbf{x}') = \sigma_f^2 \left(1 + \frac{\sqrt{3}\,d}{\ell}\right)
\exp\left(-\frac{\sqrt{3}\,d}{\ell}\right),
\label{eq:matern}
\end{equation}
where $d = \|\mathbf{x} - \mathbf{x}'\|$ is the Euclidean distance in SOAP
descriptor space, $\ell$ is the characteristic length scale, and $\sigma_f^2$
is the signal variance. The choice of $\nu = 3/2$ (gives once-differentiable functions) encodes the prior that
confinement parameters vary smoothly but not necessarily infinitely
differentiably with the local environment, consistent with the physical
analysis presented in Appendix~\ref{main-app:linear} of the main text. The hyperparameters
$\sigma_f$ and $\ell$ are optimized by maximizing the log marginal
likelihood using the L-BFGS-B algorithm with 10 random restarts. Target
values are not normalized prior to training.

The optimized hyperparameters for all models are reported in Table~\ref{supptab:gpr_hyperparameters}.
The dominant confinement onset parameter $r_{0,\mathrm{Ni}3d}$ exhibits the
longest length scale ($\ell = 0.633$), indicating the smoothest variation
across descriptor space and confirming that the physically most important
parameter is also the most learnable. The independent treatment of all
models is justified by negligible cross-parameter and cross-shell
correlations (Pearson $|r| < 0.25$ in all cases).

\begin{figure}
    \centering
    \includegraphics[width=1\linewidth]{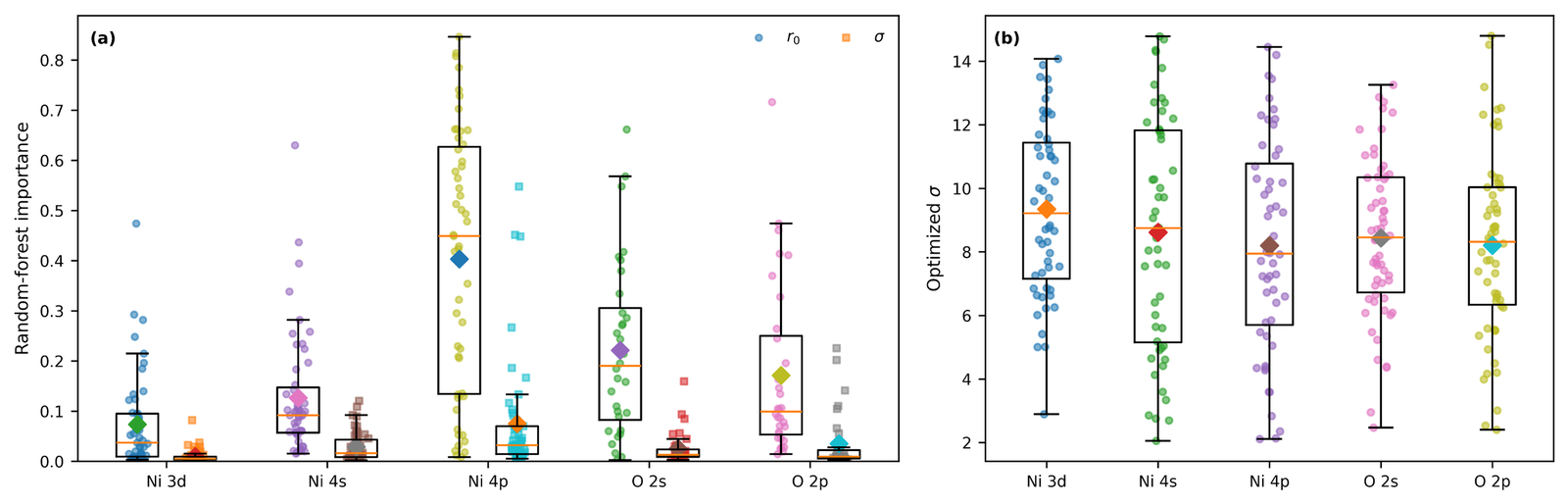}
   \caption{Confinement-parameter importance and optimized decay exponents. (a) Random-forest importance of $r_0$ and $\sigma$, obtained from the sampled PSO parameter vectors and band-structure fitting costs. Importances are normalized over the full parameter set within each optimization run and summarized by shell across runs; $r_0$ and $\sigma$ are compared within the same shell. (b) Distributions of optimized $\sigma$ values. The mean for each shell over the training set gives the fixed exponent listed in Table~\ref{supptab:gpr_hyperparameters}.}
\label{suppfig:confinement_importance}
\end{figure}

\begin{table}[h]
\caption{GPR hyperparameters for the five environment-dependent $r_0$ models and the fixed shell-dependent decay exponents $\sigma$ used in DOVE. Each fixed $\sigma$ is the mean optimized value for that shell over the training set. $\sigma_f$ denotes the GPR signal amplitude and $\ell$ the characteristic length scale of the Mat\'ern $\nu=3/2$ kernel.}
\label{supptab:gpr_hyperparameters}
\begin{ruledtabular}
\begin{tabular}{lccc}
Shell & $\sigma_{f,r_0}$ & $\ell_{r_0}$ & fixed $\sigma$ \\
\hline
Ni $3d$ & 2.29 & 0.633 & 9.3454 \\
Ni $4s$ & 2.05 & 0.447 & 8.6165 \\
Ni $4p$ & 2.10 & 0.497 & 8.1973 \\
O $2s$  & 1.56 & 0.224 & 8.4397 \\
O $2p$  & 2.48 & 0.327 & 8.2097 \\
\end{tabular}
\end{ruledtabular}
\end{table}
The optimized confinement parameters provide shell-resolved labels for both the onset radius $r_0$ and the decay exponent $\sigma$. To assess their relative importance, we fit a random-forest regressor for each PSO run, using the sampled confinement-parameter vectors as inputs and the corresponding band-structure fitting cost as the target. Parameter importance is measured by the mean decrease in impurity across the trees, normalized over the full parameter set within each run and summarized across runs. As shown in Fig.~\ref{suppfig:confinement_importance}(a), $\sigma$ generally has lower importance than $r_0$ for the same shell. This supports simplifying the predictor by fixing $\sigma$ within each shell while retaining the environment-dependent onset radii. The corresponding GPR learning curves are discussed in Sec.~\ref{supp:learning}.

We test this simplification by replacing the GPR-predicted $\sigma$ values with the shell means in Table~\ref{supptab:gpr_hyperparameters}, keeping the original $r_0$ regressors and their predictions unchanged. In the paired Materials Project comparison, the mean band-energy MAE changes from 0.3516 to 0.3487~eV. Fixing $\sigma$ therefore retains essentially the same aggregate band-energy accuracy, with the environment dependence of the final DOVE model carried by $r_0$.

\subsection{Prediction and parameter assignment}

For a given target structure, DOVE proceeds as follows: (i)~SOAP descriptors
are computed for each atom; (ii)~the trained GPR models predict the onset radii $r_0$ for each atom and shell, while the decay exponents $\sigma$ are assigned the fixed shell values in Table~\ref{supptab:gpr_hyperparameters}; (iii)~per-atom Slater-Koster files are generated using {\tt SKProgs} with the predicted $r_0$ and the fixed $\sigma$; the eigenvalues and Hubbard values are also fixed; and (iv)~a DFTB calculation is performed using \texttt{dftb+} with the atom-resolved SK tables. The GPR predictions include uncertainty estimates (posterior variance), which can serve as a diagnostic for extrapolation beyond the training domain.

\subsection{Benchmarking}

The DOVE model is benchmarked on all 37 Ni--O binary compounds from the
Materials Project~\cite{material_project_Jain2013} that are not part of the training set. For each structure, band structures are computed at the DFT-PBE and DOVE-DFTB
levels and compared using the band-energy MAE and the Gaussian accuracy defined in
Appendix~\ref{main-app:gauss} of the main text. The benchmark includes comparisons against
published trans-3d parameters~\cite{mio-trans3d} (with and without DFTB3
corrections~\cite{Nickel-DFTB3}), as well as single-type parameterizations
optimized by DSKO for Ni, Ni$_2$O, and NiO individually. Results are
presented in Fig.~\ref{main-fig:SK_integral}(c) of the main text.

\subsection{\label{supp:heldout}Transferability test in held-out set}

\begin{figure}
    \centering
    \includegraphics[width=1\linewidth]{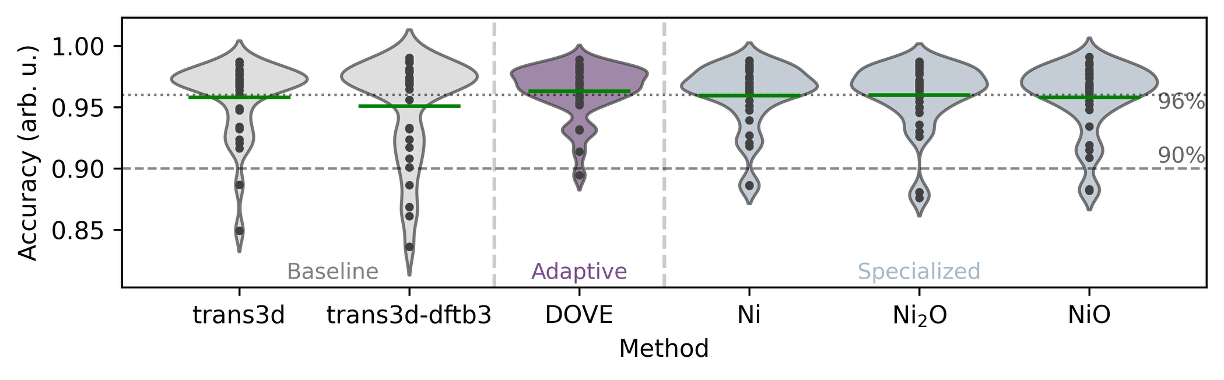}
    \caption{Gaussian accuracy for the held-out Ni-O binary structures from the evolutionary dataset, comparing baseline (trans3d, 
trans3d-dftb3), adaptive (DOVE), and material-specific (Ni, 
\ce{Ni2O}, NiO) parameterizations. The green line is the mean.}
    \label{suppfig:out-of-domain}
\end{figure}

To further assess the robustness of the DOVE model, we evaluate all
methods on a held-out set of Ni--O binary structures generated within
the same compositional workflow as the training data but not seen
during training. As shown in Fig.~\ref{suppfig:out-of-domain}, DOVE exhibits the most compact score distribution among all methods, with
the highest floor and narrowest spread. The baselines and
material-specific parameterizations show comparable median Gaussian accuracy
but wider tails, confirming that the adaptive approach provides more
consistent performance across diverse local environments.

\subsection{\label{supp:learning}Learning curve analysis}

\begin{figure}
    \centering
    \includegraphics[width=1\linewidth]{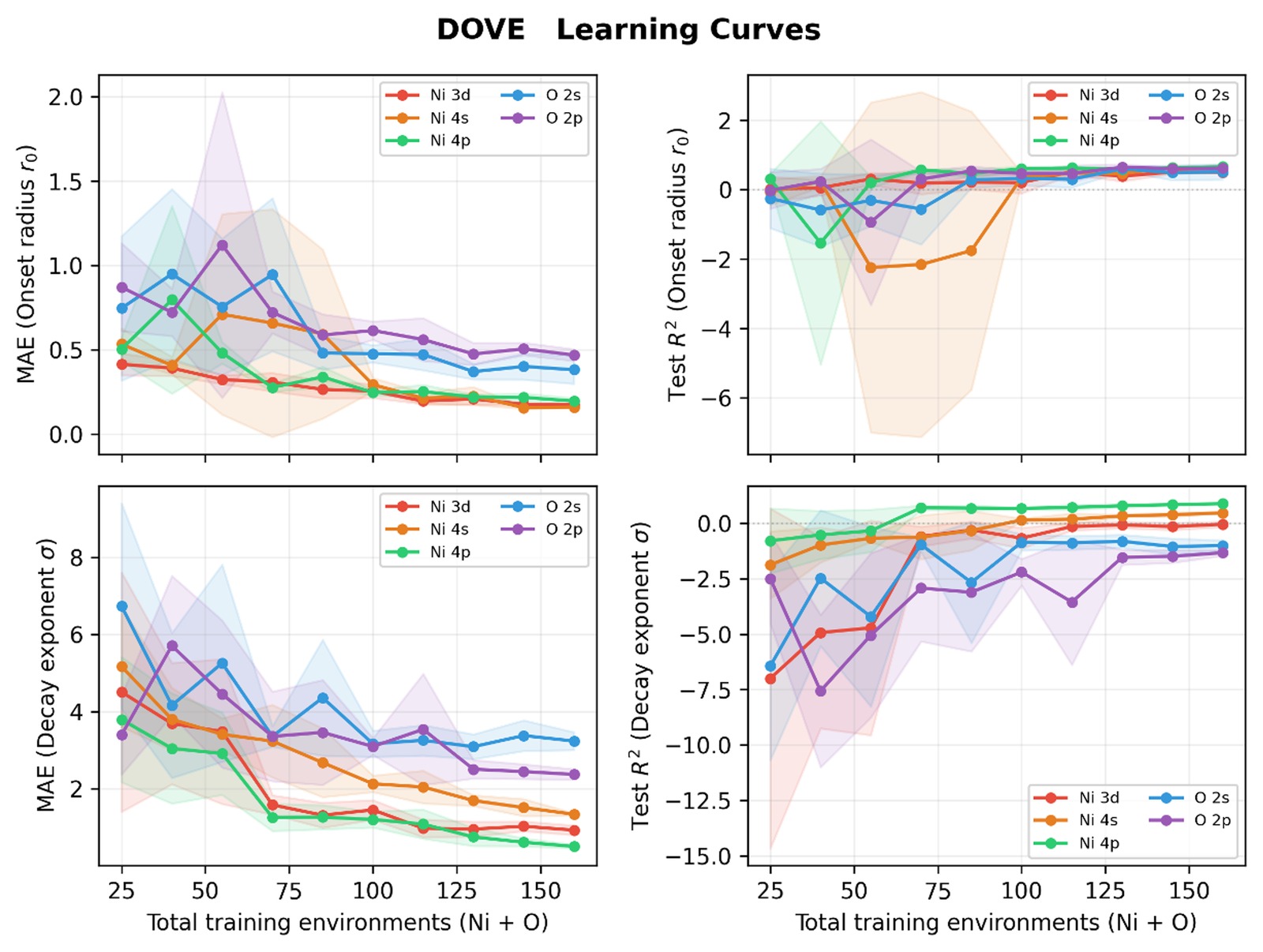}
\caption{Regression diagnostics for the optimized confinement labels: learning curves for GPR models showing prediction MAE (left) and test $R^2$ (right) for the onset radius $r_0$ (top) and the optimized decay exponent $\sigma$ (bottom). The final DOVE predictor contains GPR models for $r_0$ only. Shaded regions indicate the standard deviation over five random train/test splits with proportion 80\%/20\%.}
\label{suppfig:learning_curve}
\end{figure}
Fig.~\ref{suppfig:learning_curve} shows the GPR prediction error (MAE) and coefficient of determination ($R^2$) as a function of the total number of training environments for both confinement parameters. The $r_0$ prediction errors decrease systematically, with the dominant $r_{0,\mathrm{Ni}3d}$ reaching an MAE of approximately 0.2~Bohr and saturating near 100 environments. For the optimized $\sigma$ labels, the same GPR setup yields test $R^2$ values near or below zero in all shells, indicating substantially weaker predictive performance than for $r_0$. This motivates restricting the ML targets to $r_0$ and fixing $\sigma$ per shell (Sec.~\ref{supp:gpr}).

\subsection{Band energy MAE analysis}\label{supp:mae}

\begin{figure}[h]
\centering
\includegraphics[width=\columnwidth]{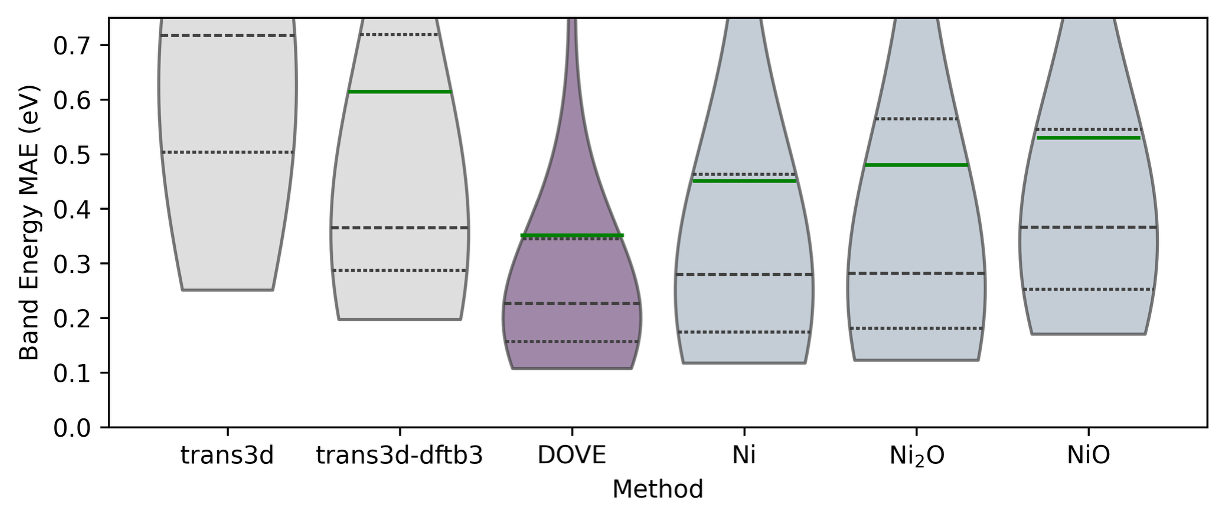}
\caption{Zoom-in Band energy MAE (eV) for the Ni-O binary structures 
from the Materials Project, comparing baseline (trans3d, 
trans3d-dftb3), adaptive (DOVE), and material-specific (Ni, 
\ce{Ni2O}, NiO) parameterizations. Green line is the mean.}
\label{suppfig:mae_violin}
\end{figure}

Fig.~\ref{suppfig:mae_violin} reports the conventional band energy MAE for the same Materials Project benchmark shown in Fig.~\ref{main-fig:SK_integral}(c) of the main text. DOVE yields the lowest median MAE among all methods, consistent with the Gaussian difference analysis. However, the MAE distributions overlap considerably across methods, illustrating the reduced sensitivity of eigenvalue-based metrics in dense band manifolds as discussed in Appendix~\ref{main-app:gauss} of the main text.

For the Materials Project benchmark, we select the contiguous DFT band block whose energy ranges overlap $[-5,+5]$ eV and compare the corresponding DFT/DFTB eigenvalues at all sampled $k$-points after subtracting each spectrum's stored energy reference. The MAE is computed per structure and then averaged over structures.

\subsection{\label{supp:accel}Acceleration of on-the-fly SK integral generation}

In the adaptive scheme, SK integrals must be regenerated whenever DOVE predicts new confinement parameters. To minimize this overhead, three optimizations are employed. First, integrals are computed only over the interatomic distance range relevant to the target structure, rather than the full tabulation grid. Second, previously computed SK integral surfaces are cached and reused: when a predicted confinement parameter falls within an already-sampled region, the corresponding integrals are interpolated rather than recomputed. Third, the updated integral values are written directly to the calculator's internal memory via a Fortran--C--Python interface, bypassing file I/O—an infrastructure that also lays the groundwork for future two-center models that predict SK integrals directly.

\bibliographystyle{apsrev4-2}

\providecommand{\TODO}[1]{{\color{red}\bfseries[TODO: #1]}}

\section{\label{supp:coo}\ce{CoO_x}}

\subsection{\label{supp:coo_refs}Reference systems and parameterization}

DFT reference band structures were computed with the settings of Sec.~\ref{sec:Computational Detail}. As for \ce{NiO_x}, all references are non-spin-polarized, for consistency with the Ni--O parameterization; the transferability of confinements optimized without spin polarization is discussed for NiO in Sec.~\ref{supp:spin}.

Two series of reference systems were considered. (i) An oxidation series built on the same parent fcc framework, in analogy to \ce{NiO_x}: bulk Co, \ce{Co2O} constructed by removing one of the two O atoms from a $(2\times1\times1)$ supercell of the rocksalt CoO primitive cell while retaining the fcc Co sublattice, and rocksalt CoO. (ii) A coordination series at fixed or varying oxidation state: zincblende CoO ($F\bar{4}3m$, parent mp-22408) for tetrahedral Co(II), rocksalt CoO ($Fm\bar{3}m$, parent mp-19079) for octahedral Co(II), and an ordered-vacancy \ce{Co2O3} structure ($P\bar{3}m1$) derived from a threefold rocksalt supercell by removing one of the three Co atoms for octahedral Co(III).

All parameterizations were performed with DSKO using PSO with 72 particles and 80 iterations, optimizing only the onset radii $r_0$ of the power-law confinement. The shell-dependent confinement exponents $\sigma$ were kept fixed to the values in Table~\ref{supptab:co_sigma} for all Co reference systems.

\begin{table}[h]
\centering
\caption{Fixed shell-dependent confinement exponents $\sigma$ used for the Co--O reference systems. Each value is the arithmetic mean of the corresponding optimized exponents from 15 preliminary Co--O parameterizations in which $\sigma$ was allowed to vary.}
\label{supptab:co_sigma}
\begin{ruledtabular}
\begin{tabular}{lccccc}
Shell & Co $3d$ & Co $4s$ & Co $4p$ & O $2s$ & O $2p$ \\
\hline
$\sigma$ & 9.7468& 7.9879& 8.6871& 9.0155& 9.2142\\
\end{tabular}
\end{ruledtabular}

\end{table}

\subsection{\label{supp:coo_r0}Optimal onset radii}

Table~\ref{supptab:co_r0} lists the optimal $r_{0,\mathrm{Co}3d}$ for all reference systems and distances, corresponding to the data points in Fig.~\ref{main-fig:apx-2}(b,c) of the main text, together with the least-squares coefficients of $r_0 = a d + b$.

\begin{table*}[t]
\centering
\caption{Optimal $r_{0,\mathrm{Co}3d}$ versus nearest-neighbor distance $d$ ($d_{\mathrm{Co-Co}}$ for the fcc-derived series, $d_{\mathrm{Co-O}}$ for the coordination series), and coefficients of the linear fits shown in Fig.~\ref{main-fig:apx-2} of the main text. The rocksalt CoO and $O_h$ Co(II) rows refer to the same calculations, expressed in terms of $d_{\mathrm{Co-Co}}$ and $d_{\mathrm{Co-O}} = d_{\mathrm{Co-Co}}/\sqrt{2}$, respectively.}
\label{supptab:co_r0}
\begin{ruledtabular}
\begin{tabular}{lcccc}
System & $d$ (\AA) & $r_{0,\mathrm{Co}3d}$ (Bohr) & $a$ (Bohr/\AA) & $b$ (Bohr) \\
\hline
Co (fcc) &
2.687, 2.899, 3.111 &
3.688, 3.912, 4.146 &
1.081 & 0.781 \\

\ce{Co2O} &
2.687, 2.899, 3.111 &
3.975, 4.450, 4.892 &
2.160 & $-$1.824 \\

CoO &
2.687, 2.899, 3.111 &
3.443, 4.058, 4.597 &
2.721 & $-$3.857 \\
\hline
$T_d$ Co(II) &
1.850, 2.000, 2.150 &
4.435, 4.877, 5.561 &
3.755 & $-$2.551 \\

$O_h$ Co(II) &
1.900, 2.050, 2.200 &
3.443, 4.058, 4.597 &
3.848 & $-$3.857 \\

$O_h$ Co(III) &
1.900, 2.050, 2.200 &
3.127, 3.591, 4.386 &
4.198 & $-$4.904 \\
\end{tabular}
\end{ruledtabular}

\end{table*}


\subsection{\label{supp:co3o4}Site-resolved prediction for \ce{Co3O4}}

\begin{figure}[h]
\centering
\includegraphics[width=1\linewidth]{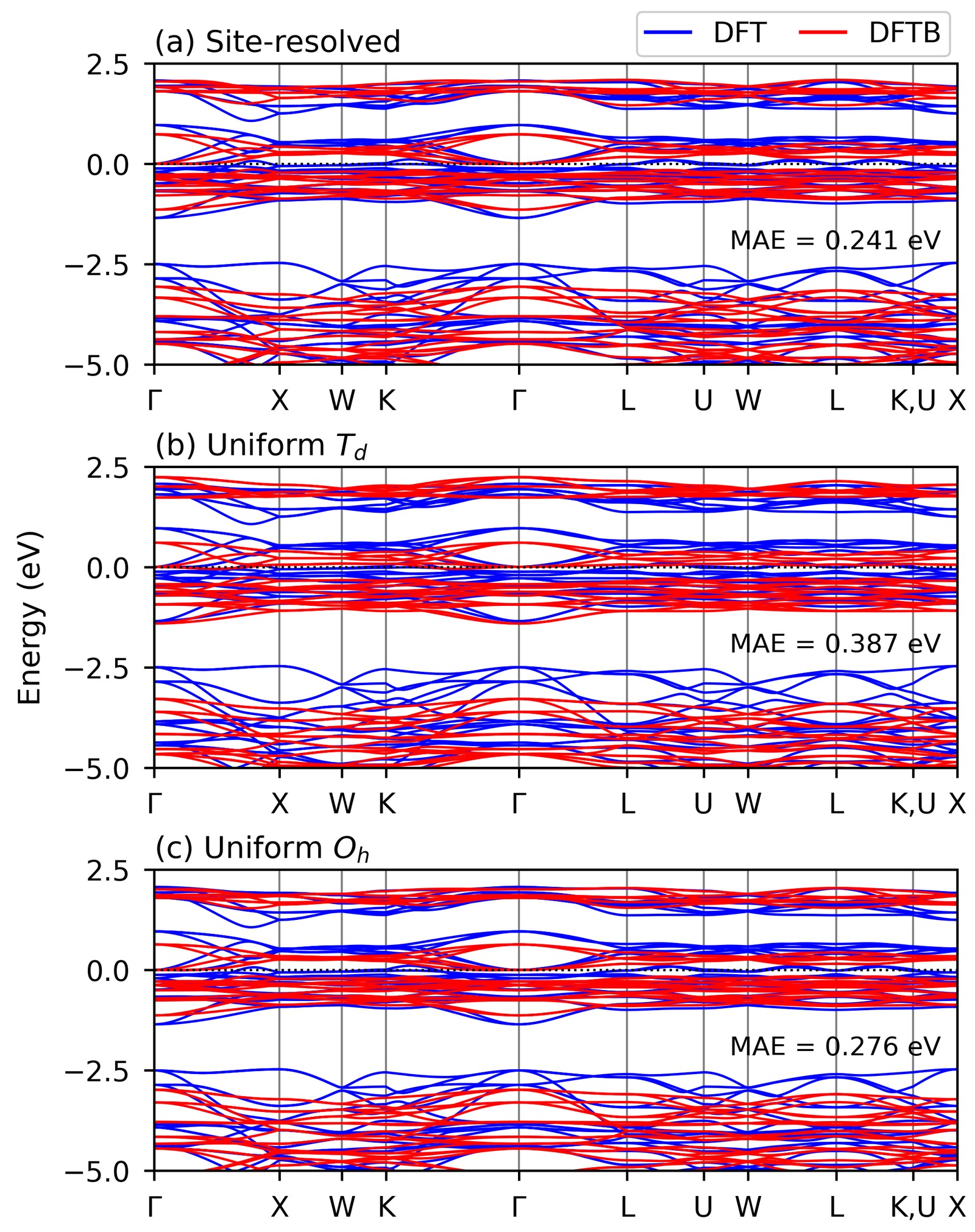}
\caption{Band structures of \ce{Co3O4} obtained with the DFT reference (blue) and DFTB (red) using (a) site-resolved, (b) uniform $T_d$-derived, and (c) uniform $O_h$-derived Co parameters. To emphasize the bands near the Fermi level, only the $-5$ to $+2.5$~eV energy range is shown. The reported MAE values are instead evaluated over the full $-15$ to $+10$~eV comparison window, yielding 0.241, 0.387, and 0.276~eV for the site-resolved, uniform $T_d$, and uniform $O_h$ assignments, respectively.}
\label{suppfig:co3o4_bands}
\end{figure}

The spinel \ce{Co3O4} structure was taken from mp-18748, retained without further relaxation. Its primitive cell contains 2 tetrahedral Co(II) and 4 octahedral Co(III) sites, with $d_{\mathrm{Co-O}} = $ 1.955 \AA{} and 2.025 \AA{}, respectively. For each site, $r_{0,\mathrm{Co}3d}$ was obtained by evaluating the linear fit of the corresponding motif (Table~\ref{supptab:co_r0}) at this distance, without any \ce{Co3O4} electronic-structure information. In the two uniform assignments, the $r_0$ derived from the corresponding $T_d$ and $O_h$ fit, respectively, is applied to all Co atoms. The resulting parameters and band-energy MAEs are listed in Table~\ref{supptab:co3o4}, and the corresponding band structures are compared in Fig.~\ref{suppfig:co3o4_bands}.

Band-energy MAEs were evaluated over $-15$ to $+10$~eV, following the procedure of Sec.~\ref{supp:mae}.

\begin{table}[h]
\centering
\caption{Onset radii $r_0^{T_d}$ and $r_0^{O_h}$ (Bohr) assigned to the tetrahedral and octahedral Co sites of \ce{Co3O4}, and resulting band-energy MAE with respect to DFT ($-15$ to $+10$~eV).}
\label{supptab:co3o4}
\begin{ruledtabular}
\begin{tabular}{lccc}
Assignment & $r_0^{T_d}$ & $r_0^{O_h}$ & MAE (eV) \\
\hline
Site-resolved & 4.787& 3.595& 0.241 \\
Uniform, $T_d$-derived & 4.787& 4.787& 0.387 \\
Uniform, $O_h$-derived & 3.595& 3.595& 0.276 \\
\end{tabular}
\end{ruledtabular}

\end{table}

\bibliography{references}